\documentclass[floats,floatfix,amssymb,prd,twocolumn,superscriptaddress,nofootinbib]{revtex4-1}

\usepackage{subcaption}
\usepackage{ragged2e}
\DeclareCaptionJustification{justified}{\justifying}
\makeatletter
\newcommand{\subsetsim}{\mathrel{\mathpalette\subset@sim\relax}}
\newcommand{\subset@sim}[2]{%
  \vtop{\offinterlineskip\m@th
    \ialign{\hfil##\cr
      $#1\subset$\cr\noalign{\kern0.5pt}\scalebox{0.9}{$#1\sim$}\cr
    }%
  }%
}
\makeatother

\usepackage{amssymb,amsmath,verbatim,mathtools,needspace,enumitem,etoolbox,graphicx,physics,microtype,afterpage,bm}

\usepackage[dvipsnames, usenames]{xcolor}
\definecolor{linkcolor}{rgb}{0.0,0.3,0.5}
\usepackage{booktabs}

\definecolor{oucrimsonred}{rgb}{0.6, 0.0, 0.0}
\definecolor{persianblue}{rgb}{0.11, 0.22, 0.73}
\definecolor{forestgreen}{rgb}{0.13,0.35,0.13}

\usepackage[unicode, 
colorlinks=true, 
linkcolor=persianblue, 
citecolor=forestgreen, 
filecolor=persianblue,
urlcolor=oucrimsonred, 
pdfusetitle]{hyperref}

\usepackage[all]{hypcap}
\usepackage[T1]{fontenc}
\usepackage[utf8]{inputenc}
\usepackage{tabularx}
\usepackage{adjustbox}
\usepackage{float}
\usepackage{ulem}
\usepackage{amsmath}
\usepackage{xfrac}
\usepackage{orcidlink}

\definecolor{azure}{rgb}{0.0, 0.5, 1.0}
\definecolor{deepfuchsia}{rgb}{0.76, 0.33, 0.76}
\definecolor{VioletRed4}{rgb}{0.55, 0.13, .32}
 
\definecolor{harvardcrimson}{rgb}{0.79, 0.0, 0.09}
\definecolor{oceanboatblue}{rgb}{0.0, 0.47, 0.75}
\definecolor{persianblue}{rgb}{0.11, 0.22, 0.73}
\definecolor{egyptianblue}{rgb}{0.06, 0.2, 0.65}
\definecolor{navyblue}{rgb}{0.0, 0.0, 0.5}
\newcommand{\td}{{\rm d}}
\usepackage{multirow}
\usepackage{pifont}
\usepackage{fontawesome}
\usepackage{lmodern}
\usetikzlibrary{arrows.meta}

\usepackage{multirow}

\allowdisplaybreaks
\usepackage{tikz}
\usepackage{tcolorbox}
\usetikzlibrary{tikzmark}
\usepackage{color}
\usepackage{xcolor}
\usepackage{framed}

\definecolor{rossos}{cmyk}{0,1,1,0.55}
\definecolor{bluscuro}{rgb}{0.15, 0.2, .85}
\definecolor{bluchiaro}{cmyk}{1,.3,0.,0.1}
\definecolor{ForestGreen}{rgb}{0.13, 0.55, 0.13}

\definecolor{rossocorsa}{rgb}{0.9, 0.0, 0.0}

\newcommand{\MPl}{\bar{M}_{\textrm{\tiny{Pl}}}}

\def\nn{\nonumber}

\def\bea{\begin{eqnarray}}
\def\eea{\end{eqnarray}}

\newcommand{\bs}{\begin{subequations}}
\newcommand{\es}{\end{subequations}}

\newcommand{\be}{\begin{equation}}
\newcommand{\ee}{\end{equation}}

\def\lsim{\mathrel{\rlap{\lower4pt\hbox{\hskip0.5pt$\sim$}}
    \raise1pt\hbox{$<$}}}         
\def\gsim{\mathrel{\rlap{\lower4pt\hbox{\hskip0.5pt$\sim$}}
    \raise1pt\hbox{$>$}}}         

\makeatletter
\def\l@subsubsection#1#2{}
\makeatother

\begin{document}
\title{Is the formation of 
primordial black holes from single-field inflation 
\\compatible with standard cosmology?}

\author{Sasha Allegrini\orcidlink{0009-0004-2664-7440}}
\thanks{{\scriptsize Email}: \href{mailto:allegrini.2082804@studenti.uniroma1.it}{allegrini.2082804@studenti.uniroma1.it}}
\affiliation{Dipartimento di Fisica, ``Sapienza'' Universit\`a di Roma, Piazzale Aldo Moro 5, 00185, Roma, Italy}

\author{Loris Del Grosso\orcidlink{0000-0002-6722-4629}}
\thanks{{\scriptsize Email}: \href{mailto:loris.delgrosso@uniroma1.it}{loris.delgrosso@uniroma1.it}}
\affiliation{Dipartimento di Fisica, ``Sapienza'' Universit\`a di Roma, Piazzale Aldo Moro 5, 00185, Roma, Italy}
\affiliation{INFN sezione di Roma, Piazzale Aldo Moro 5, 00185, Roma, Italy}
\affiliation{William H. Miller III Department of Physics and Astronomy,
Johns Hopkins University, Baltimore, Maryland 21218, USA}

\author{Antonio J. Iovino\orcidlink{0000-0002-8531-5962}}
\thanks{{\scriptsize Email}: \href{mailto:antoniojunior.iovino@uniroma1.it}{antoniojunior.iovino@uniroma1.it}}
\affiliation{Dipartimento di Fisica, ``Sapienza'' Universit\`a di Roma, Piazzale Aldo Moro 5, 00185, Roma, Italy}
\affiliation{INFN sezione di Roma, Piazzale Aldo Moro 5, 00185, Roma, Italy}
\affiliation{Department of Theoretical Physics and Gravitational Wave Science Center,  \\
24 quai E. Ansermet, CH-1211 Geneva 4, Switzerland}

\author{Alfredo Urbano\orcidlink{0000-0002-0488-3256}}
\thanks{{\scriptsize Email}: \href{mailto:alfredo.urbano@uniroma1.it}{alfredo.urbano@uniroma1.it}}
\affiliation{Dipartimento di Fisica, ``Sapienza'' Universit\`a di Roma, Piazzale Aldo Moro 5, 00185, Roma, Italy}
\affiliation{INFN sezione di Roma, Piazzale Aldo Moro 5, 00185, Roma, Italy}

\date{\today}

\begin{abstract}\noindent 
In this work, we investigate the generation of primordial black holes (PBHs) within the framework of single-field inflationary models and their compatibility with the cosmological history of the Universe. Our results suggest that, depending on the masses of the formed PBHs, single-field inflation models require more than fine-tuning a potential to induce ultra-slow roll; it necessitates a comprehensive understanding of the post-inflationary cosmological evolution. As an explicative example, we introduce a new model, based on a double inflection point and consistent with Cosmic Microwave Background observations, capable of generating sub-solar PBHs, whose merger could be potentially detectable by the LVK experiment.
\end{abstract}

\maketitle

\noindent
\section{Introduction and motivations}

The exploration of inflationary models that lead to the formation of primordial black holes (PBHs)\,\cite{Zeldovich:1967lct,Hawking:1974rv,Chapline:1975ojl,Carr:1975qj} is driven by a profound quest to unravel the mysteries of the early universe and its subsequent evolution. 

Inflationary models, crafted to tackle fundamental cosmological puzzles such as the horizon and flatness problems, furnish a theoretical framework for understanding the genesis of large-scale structures. In this context, PBHs, originating from the gravitational collapse of rare but highly dense regions of space-time in the early universe, emerge as captivating tools for delving into the universe's infancy at scales much shorter than those probed by Cosmic Microwave Background (CMB) measurements.

In this work, we investigate whether the formation of PBHs from single-field inflationary models is compatible with Standard Cosmology. 
Specifically, we consider PBHs that form from large spikes in the power spectrum of primordial curvature perturbations, leading to overdensities that may collapse into PBHs.
At first glance, the answer to the aforementioned question would seem to be a resounding yes. After all, as just mentioned, the formation of PBHs occurs on scales much smaller than that of the CMB, where the observational constraints on the power spectrum of curvature perturbations are extremely weak and, in fact, arise almost exclusively from the requirement to avoid overproducing PBHs\,\cite{Cole:2017gle,Gow:2020bzo,Iovino:2024tyg}.

However, in this work, we aim to demonstrate how accommodating the production of PBHs in the evolutionary history of the universe disrupts a delicate balance of scales, potentially conflicting with standard evolutionary dynamics.

Firstly, let us clearly define the context we have in mind.
While traditional Standard Cosmology focuses on the Big Bang and subsequent expansion, Modern Cosmology often incorporates inflation and reheating as integral parts of the overall picture. 
Reheating marks the end of the inflationary phase and the beginning of the hot Big Bang phase, setting the stage for the standard cosmological evolution, including nucleosynthesis and the formation of the CMB.
The perspective we choose to adopt is precisely this one. 
In particular, reheating is crucial because it connects the end of inflation to the well-established hot Big Bang model, ensuring a smooth transition and explaining the thermalization of the universe.

Various reheating mechanisms have been extensively discussed in the literature. In perturbative reheating\,\cite{Dolgov:1982th,Abbott:1982hn,Dolgov:1989us,Shtanov:1994ce,Chung:1998rq,Kolb:2003ke}, after inflation ends, the inflaton field oscillates around the minimum of its potential. These oscillations cause the inflaton to decay into Standard Model particles via perturbative processes, gradually heating the universe. In this case, reheating is a slow process with a gradual transfer of energy, governed by the decay rate of the inflaton.
On the other hand, the efficiency of the reheating process is greatly enhanced in the presence of a preheating stage\,\cite{Kofman:1994rk,Kofman:1997yn,Greene:1997fu,Felder:1998vq}. Preheating is characterized by non-perturbative effects that lead to a rapid and explosive production of particles. In this case, resonance effects cause the inflaton field to transfer energy to other fields very efficiently.  

The reheating epoch can be characterized by an equation of state parameter, denoted by $\omega_{\textrm{reh}}$, which is commonly defined as the ratio of pressure to energy density. During different stages of reheating, the effective equation of state parameter can change, reflecting the underlying dynamics of the processes involved.
However, we can draw some general considerations.

When the inflation era ends, the equation of state parameter of the inflaton field equals $-1/3$. On the other hand, the radiation era is characterized by an equation of state parameter equal to $+1/3$. It is therefore natural to postulate that the reheating stage interpolates between these two cases, 
$-1/3 \leqslant \omega_{\textrm{reh}} \leqslant 1/3$. We refer to this situation as standard reheating.

Schematically, and without any claim to formality, the cosmological evolution we envision proceeds through the following steps
\begin{align}
	\begin{tikzpicture}
	 {\scalebox{1}{
    \node at (-3,0.75) {\scalebox{0.95}{$\textrm{single-field}$}};
    \node at (-3,0.4) {\scalebox{0.95}{$\textrm{inflation}$}};
    \node at (-1,0.75) {\scalebox{0.95}{$\textrm{reheating}$}};
    \node at (-1,0.4) {\scalebox{0.95}{$\textrm{$-\frac{1}{3}\leqslant \omega_{\textrm{reh}}
\leqslant \frac{1}{3}
$}$}};
    \node at (1.75,0.75) {\scalebox{0.95}{$\textrm{standard Big Bang}$}};
    \node at (1.75,0.4) {\scalebox{0.95}{$\textrm{cosmology}$}};
    \node at (4,0.575) {\scalebox{0.95}{$\textrm{today}$}};
    \draw[->,>=Latex,thick] (-3.5,0)--(4.0,0);
     \node at (0.85,-0.35) {\scalebox{0.95}{
     {\color{oucrimsonred}{$\textrm{PBH formation?}$}}}};
    }}
	\end{tikzpicture}\nn
\end{align} 

This work is organized as follows. 
In section\,\ref{sec:ScalesUniverse},  we elaborate more formally on the evolution of the universe sketched in the previous schematic. Many of the topics we will discuss in this section are not original material. However, we believe that it is crucial to explain them in the necessary detail because they contain all the physics needed to understand the main point of our work.
In section\,\ref{sec:PBHprob}, with the help of a toy model, we discuss the issues of the mismatch of scales in the presence of a period of ultra-slow roll, capable of producing a large amount of PBH.
In section\,\ref{sec:Model}, in order to demonstrate further a concrete application of our ideas, we introduce an explicit model of single-field dynamics, based on a double inflection point, capable of produce sub-solar mass PBHs. We conclude in section\,\ref{sec:Conc}.

\noindent
\section{The Timeline of the universe}\label{sec:ScalesUniverse}
Cosmology is a tale of scales, and to fully understand the universe, we must study and integrate knowledge across an enormous range of scales, from the quantum realm to the cosmic horizon. Each scale contributes to our understanding of how the universe began, evolved, and continues to change over time.

We focus our attention on the time evolution of the comoving Hubble radius $R_H\equiv (aH)^{-1}$. In this expression, $a=a(t)$ is the scale factor of the flat Friedmann-Robertson-Walker (FRW) metric whose line element is commonly expressed as $ds^2 = dt^2 - a(t)^2[d\chi^2+ \chi^2(d\theta^2+sin^2\theta d\phi^2)]$ with $t$ the cosmic time and $(\chi,\theta,\phi)$ the comoving spherical coordinates.
The Hubble expansion rate is defined by $H \equiv \dot{a}/a$ where the overdot denotes a derivative with respect to cosmic time, that is, $\dot{a}=da/dt$. 
We use subscripts $_{0}$ to denote the quantities evaluated today, at $t = t_0$. 
In this section, we employ the standard normalization for the scale factor, assuming $a_0 = 1$.
We introduce the conformal time $\tau$ defined by $d\tau = dt/a(t)$. 

An important point for our discussion is that inside the comoving Hubble radius, regions are causally connected, meaning that they can influence each other via light signals or other causal interactions. Outside this radius, regions are receding from each other faster than the speed of light because of the expansion of the universe, making them causally disconnected.

The aim now is to utilize this information to extract insight and potentially impose constraints on inflationary models\,\cite{Liddle:2003as} capable of produce a substantially amount of PBHs. 
Inflation provides a compelling solution to the horizon problem in cosmology. 
The horizon problem arises from the challenge of explaining the observed isotropy of the CMB radiation. The latter exhibits an almost perfect isotropy, with temperature anisotropies smaller than one part per ten thousand.
In the standard big bang cosmology, the universe underwent rapid expansion during its early phases (driven first by radiation then by matter). This expansion led to non-overlapping past light cones for many regions in the observable universe, implying that these regions never had the opportunity for direct causal contact. According to conventional cosmological principles, such regions, widely separated and never in causal contact, should not exhibit similar temperatures. 

For a universe dominated by a fluid with equation of state
$P = \omega \rho$, the evolution of the comoving Hubble radius is dictated by 
\begin{align}
\frac{1}{aH} \propto a^{(1+3\omega)/2}\,.\label{eq:PerfectFluid}
\end{align}
Inflation addresses the horizon problem by proposing a phase of decreasing comoving Hubble radius in the early universe, where $\frac{d}{dt}(1/aH) < 0$. Eq.\,(\ref{eq:PerfectFluid}) thus implies that one needs $\omega < -1/3$. 
The inflaton field is a hypothetical scalar field $\phi$ which is conjectured to have driven cosmic inflation in the very early universe\,\cite{Guth:1980zm,Linde:1981mu}. 
This is because pressure and energy density are given by 
\begin{align}
P_{\phi} = \frac{1}{2}\dot{\phi}^2 - V(\phi)\,,~~~
\rho_{\phi} = \frac{1}{2}\dot{\phi}^2 + 
V(\phi)\,,\label{eq:InflaField}
\end{align}
where consistency with the symmetries of the FRW spacetime requires that the value of the inflaton only depends on time, $\phi=\phi(t)$. Under the assumption that the potential term dominates over the kinetic energy one gets $P_{\phi} \approx -\rho_{\phi}$.
From Eq.\,(\ref{eq:PerfectFluid}) with $\omega = -1$, we find that the Hubble rate remains approximately constant during inflation, while the scale factor undergoes exponential expansion, $a(t) \propto e^{Ht}$. 

The transition between the inflationary phase and the radiation era is referred to as the reheating phase. 
The mechanism through which reheating occurs depends on the specific details of the particle physics involved in the early universe. During reheating, the inflaton field oscillates around its minimum and its energy is transferred to other particles, leading to the creation of a hot and dense bath of particles. This process is essential for bridging the inflationary phase to the subsequent radiation-dominated era, laying the groundwork for the formation of light elements and the observed evolution of the universe. Although reheating is ideally modeled as an instantaneous transition, a more realistic approach requires detailed modeling. In a more comprehensive description, reheating involves the complex process of converting the energy stored in the inflaton field into particles that make up the Standard Model of particle physics. Traditional analyses of reheating focus on the dynamics of inflaton decay, particle production, and thermalization.

In this work, we essentially remain agnostic about the details of reheating (although in the discussion of our results we will strive to be as concrete as possible). For the time being, we will simply operate under the assumption that between the end of inflation and the onset of the radiation era, the universe undergoes a phase during which it is dominated by a fluid with a constant equation of state, $\omega_{\textrm{reh}}$.

At this point, it is useful to introduce the $e$-fold number $N$. The relation between the $e$-fold number and the cosmic time is  expressed using the equation
\begin{align}
dN = Hdt = d\log a\,.\label{eq:EfoldNumber}
\end{align}
This relation can be integrated within some time interval $[t_{\textrm{in}},
t_{\textrm{fin}}]$, and gives 
\begin{align}
N_{\textrm{fin}} - N_{\textrm{in}} = \int_{t_{\textrm{in}}}^{t_{\textrm{fin}}}Hdt = \log\left(\frac{a_{\textrm{fin}}}{a_{\textrm{in}}}\right)
\Rightarrow
\frac{a_{\textrm{fin}}}{a_{\textrm{in}}} = e^{N_{\textrm{fin}} - N_{\textrm{in}}}\,.
\end{align}
Using the number of $e$-folds as a measure of time, we will frequently employ the following shorthand notation for the time evolution of the Hubble rate and scale factor
\begin{align}
H(N_{i}) \equiv H_i\,,~~~a(N_i) \equiv a_i\,.   
\end{align}

We have now reached the position of finally discussing one of the key points of our analysis. 
We consider some value of comoving wavenumber $k$ and compare it with the inverse comoving Hubble radius at the time of matter-radiation equality, $k/a_{\textrm{eq}}H_{\textrm{eq}}$. 
We introduce the $e$-fold time $N_k$ defined through the equation
\begin{align}
k=a(N_k)H(N_k) = a_kH_k\,.    
\end{align}

Following from our previous discussion, $N_k$ indicates the moment in time during inflation when $k$ transitions from being sub-horizon to super-horizon. We thus write (see Appendix\,\ref{app:Tales} for a derivation)
\begin{align}
\log\left(
\frac{k}{a_{\textrm{eq}}H_{\textrm{eq}}}
\right) & = -\Delta N_{k} + \frac{1}{2}
\log\left(\frac{H_k}{\sqrt{3}\MPl}\right)
+ \log\left(\frac{T_0}{H_{\textrm{eq}}}\right)\nn\\
& +\frac{(3\omega_{\textrm{reh}}-1)}{4}\Delta N_{\textrm{reh}} -
\frac{1}{4}\log\left(
\frac{30}{g_{\textrm{reh}}\pi^2}
\right)
\nn\\
& -
\frac{1}{3}\log\left(
\frac{11g_{s,\textrm{reh}}}{43}
\right) - \log\left(\frac{a_{\textrm{eq}}}{a_0}\right)\,,\label{eq:MainLog}
\end{align}
where $g_{\textrm{reh}}$ and $g_{s,\textrm{reh}}$ are respectively the relativistic degrees of freedom in energy and entropy upon thermalization.
In this expression 
\begin{itemize}
\item[{\it i)}] $\Delta N_k \equiv N_{\textrm{end}}-N_k$ indicates the number of $e$-fold between the moment in time during inflation when $k$ transitions from being sub-horizon to super-horizon and the end of inflation.
\item[{\it ii)}] $\Delta N_{\textrm{reh}}\equiv N_{\textrm{reh}}-N_{\textrm{end}}$ denotes the duration of the reheating phase in terms of the number of $e$-folds.
\item[{\it iii)}] $\Delta N_{\textrm{RD}}\equiv N_{\textrm{eq}}-N_{\textrm{reh}}$  
denotes the duration of the radiation era that elapses between the end of reheating and the matter-radiation equality in terms of the number of $e$-folds. 
\end{itemize}
According to these definitions, 
$\Delta N_{k}$, $\Delta N_{\textrm{reh}}$ and $\Delta N_{\textrm{RD}}$ are positive quantities. 

\begin{figure}[!t]
	\centering
\includegraphics[width=0.495\textwidth]{ 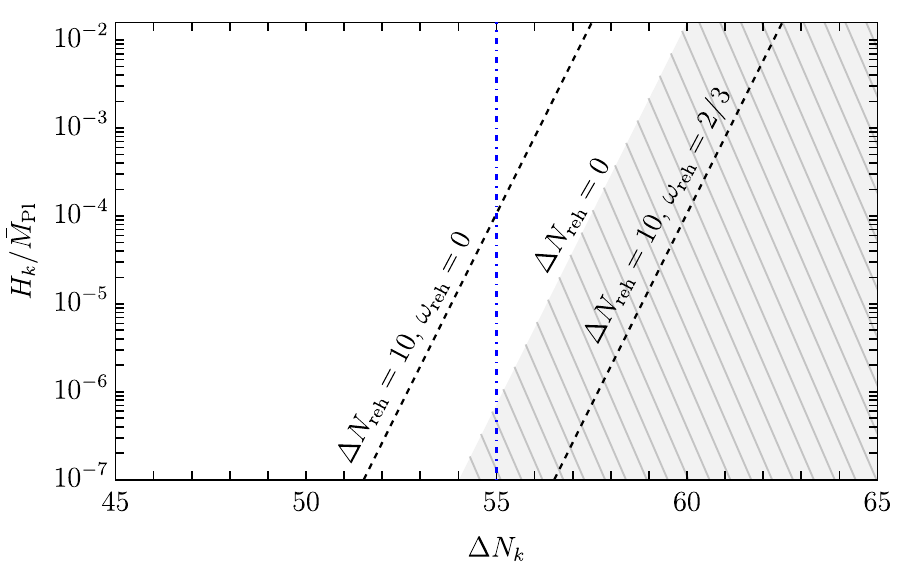}
	\caption{\it Values of the Hubble parameter $H_{k}$ when the related mode $k$ crosses the horizon for different durations of the reheating $\Delta N_{\rm reh}$ and $w_{\rm reh}$ as a function of the duration of inflation $\Delta N_k$. The region shaded in gray corresponds to the condition $\log(k/
a_{\textrm{eq}}H_{\textrm{eq}}) < 0$ computed according to Eq.\,(\ref{eq:MainLog}) assuming instantaneous reheating.
 }
\label{fig:InflaBound}
\end{figure}

Consequently, if we want the comoving wavenumber $k$ to be sub-horizon at the time of matter-radiation equality, we shall have $\log(k/a_{\textrm{eq}}H_{\textrm{eq}}) > 0$.
Through the preceding equation, this condition translates into a constraint on the parameters $\Delta N_k$, $H_k$, $\omega_{\textrm{reh}}$, and $\Delta N_{\textrm{reh}}$. 
Consider, for instance, the case of instantaneous reheating, $\Delta N_{\textrm{reh}} = 0$. 
In Fig.\,\ref{fig:InflaBound}, the gray shaded region corresponds to the condition $\log(k/a_{\textrm{eq}}H_{\textrm{eq}}) < 0$. For specificity, let's consider a mode that exits the horizon approximately $\Delta N_k = 55$ $e$-folds before the end of inflation (vertical dot-dashed blue line in Fig.\,\ref{fig:InflaBound}). If, at the time of its horizon crossing, we have $H_k/\MPl \lesssim 10^{-6}$, then the mode will still be super-horizon at the time of matter-radiation equality. Conversely, if $H_k/\MPl \gtrsim 10^{-6}$, the mode will be sub-horizon at the time of matter-radiation equality.

The inclusion of reheating changes the picture depending on its duration $\Delta N_{\textrm{reh}}$ and the sign of the factor $(3\omega_{\textrm{reh}}-1)$. 
\begin{itemize}
    \item[$\circ$] If  $\omega_{\textrm{reh}} < 1/3$, the reheating stage gives a negative contribution to the right-hand side of Eq.\,(\ref{eq:MainLog}). 
    Consequently, to keep the mode $k$ sub-horizon at the time of matter-radiation equality, one needs to compensate with a larger value of $H_k$ or a smaller $\Delta N_k$.
    \item[$\circ$] If  
    $\omega_{\textrm{reh}} > 1/3$, the reheating stage gives a positive contribution to the right-hand side of Eq.\,(\ref{eq:MainLog}). This consequently allows for the extension of the duration of the interval $\Delta N_k$ or the reduction of the value of $H_k$ without violating the condition $\log(k/a_{\textrm{eq}}H_{\textrm{eq}}) > 0$.
\end{itemize}
It should be noted that the equation of state parameter of a homogeneous condensate oscillating in a potential with a minimum of the form
$V(\phi) \propto \phi^p$ is given by
$\omega_{\textrm{reh}} = (p-2)/(p+2)$, see Ref.\,\cite{Turner:1983he}. 
Consequently, $\omega_{\textrm{reh}} > 1/3$ requires $p>4$, indicating a potential dominated near its minimum by higher-dimensional operators, a scenario that is certainly not natural. 
On the other hand, the more conventional case of a minimum dominated by the quadratic term results in $\omega_{\textrm{reh}} =0$.

It should be noted that, during reheating, the equation of state parameter $\omega_{\textrm{reh}}$ can exceed $1/3$ due to several reasons. 
For example, in kinetic dominance, where the energy of the inflaton field is mainly in the form of kinetic energy, the effective $\omega_{\textrm{reh}}$ can temporarily exceed $1/3$. 
However, it is important to stress that these scenarios typically involve transient conditions and do not represent long-term equilibrium states.

We streamline the main argument of this section with the help of a simple model of inflation. Therefore, we put aside for the moment the formation of PBHs and consider the Starobinsky model of inflation\,\cite{Starobinsky:1980te}.
The latter is based on the scalar potential 
\begin{align}
V_{\textrm{Staro}}(\phi) = 
\frac{3M^2\MPl^2}{4}
\left[1 - 
\exp\left(
-\sqrt{\frac{2}{3}}
\frac{\phi}{\MPl}
\right)
\right]^2\,.\label{eq:PoteStaro}
\end{align}
where $M$ is a fundamental mass scale that,
in the chosen normalization for the potential in Eq.\,(\ref{eq:PoteStaro}), coincides with the mass of the inflaton.
\begin{figure}[!t]
	\centering
\includegraphics[width=0.495\textwidth]{ 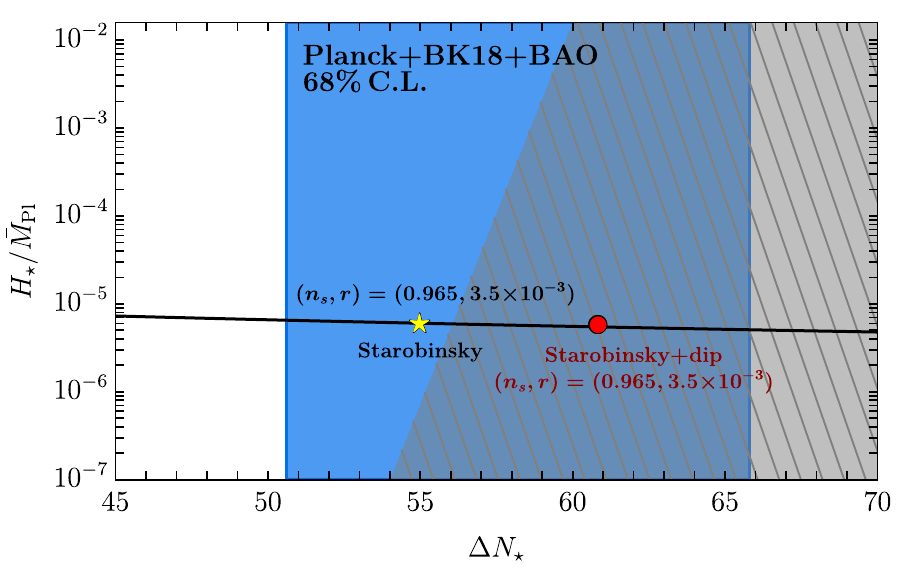}
	\caption{\it Same as Fig.\,\ref{fig:InflaBound} but focus on the CMB scale $k_*=0.05$ ${\rm Mpc}^{-1}$. The solid black line corresponds to Eq.\,(\ref{eq:HStarStaro}) with $A_s = 2.1\times 10^{-9}$. The region shaded in blue corresponds to the 68\% C.L. contour on $(n_s,r)$ using the Planck
2018 data, Ref.\,\cite{BICEP:2021xfz}. For fixed inflationary parameters $n_s = 0.965$ and $r = 3.5 \times 10^{-3}$, the yellow star and the red circle represent respectively the duration of the inflation in the Starobinsky model without and with a dip.}
\label{fig:StaroB}
\end{figure}
In the slow-roll approximation, we find (cf. appendix\,\ref{app:Staro}) 
\begin{align}
H_{\star}^2 = \frac{M^2}{4}
\left[
1+\frac{1}{W_{-1}(f_{\Delta N_{\star}})}
\right]^2\,,\label{eq:HStarStaro}
\end{align}
where $W_{-1}(z)$ is the branch with 
$k=-1$ of the Lambert W function $W_k(z)$ and $f_{\Delta N_{\star}}$ is defined in Eq.\,(\ref{eq:ShortHandfx}). 
The mass scale $M$ is fixed by the amplitude of the scalar power spectrum measured at the CMB pivot scale. We find
\begin{align}
A_s = \frac{3M^2[1+W_{-1}(f_{\Delta N_{\star}})]^4}{128\pi^2 \MPl^2 
W_{-1}(f_{\Delta N_{\star}})^2}\,.
\end{align}
On the other hand, the amount of inflation $\Delta N_{\star}$ must be compatible with 
the constraints on the scalar spectral index $n_s$ and the tensor-to-scalar ratio $r$ since these can be expressed as
\begin{align}
n_s & = 1-\frac{16}{3[1+W_{-1}(f_{\Delta N_{\star}})]^2} 
+ \frac{8}{
3[1+W_{-1}(f_{\Delta N_{\star}})]
}\,,\nn\\
r & = 
\frac{64}{
3[1+W_{-1}(f_{\Delta N_{\star}})]^2
}\,.
\end{align}
We summarize our result in Fig.\,\ref{fig:StaroB}. This is the same plot shown in Fig.\,\ref{fig:InflaBound} but now specified for the CMB pivot scale $k_{\star}$. 
The solid black line corresponds to Eq.\,(\ref{eq:HStarStaro}) with $A_s = 2.1\times 10^{-9}$. 
The region shaded in blue corresponds to the 68\% C.L. contour on $(n_s,r)$ that we obtain using the Planck
2018 baseline analysis including BICEP/Keck and BAO data (BK18, for short, in the rest of this work), cf. Ref.\,\cite{BICEP:2021xfz} (see also \href{http://bicepkeck.org/bk18_2021_release.html}{BK18 Data Products}). 
The region shaded in gray corresponds to the condition $\log(k_{\star}/
a_{\textrm{eq}}H_{\textrm{eq}}) < 0$ computed according to Eq.\,(\ref{eq:MainLog}) assuming instantaneous reheating. 

The upshot of the analysis is that it is possible to appropriately choose the value of $\Delta N_{\star}$ in order to satisfy the cosmological constraints from Planck+BICEP/Keck without violating the condition that small scales re-enter the horizon before matter-radiation equality. 
For illustration, the yellow star in Fig.\,\ref{fig:StaroB} corresponds to $\Delta N_{\star} = 55$. As evident from the plot, this inflationary solution does not create any tension. 
\begin{figure}[!t]
	\centering
\includegraphics[width=0.495\textwidth]{ 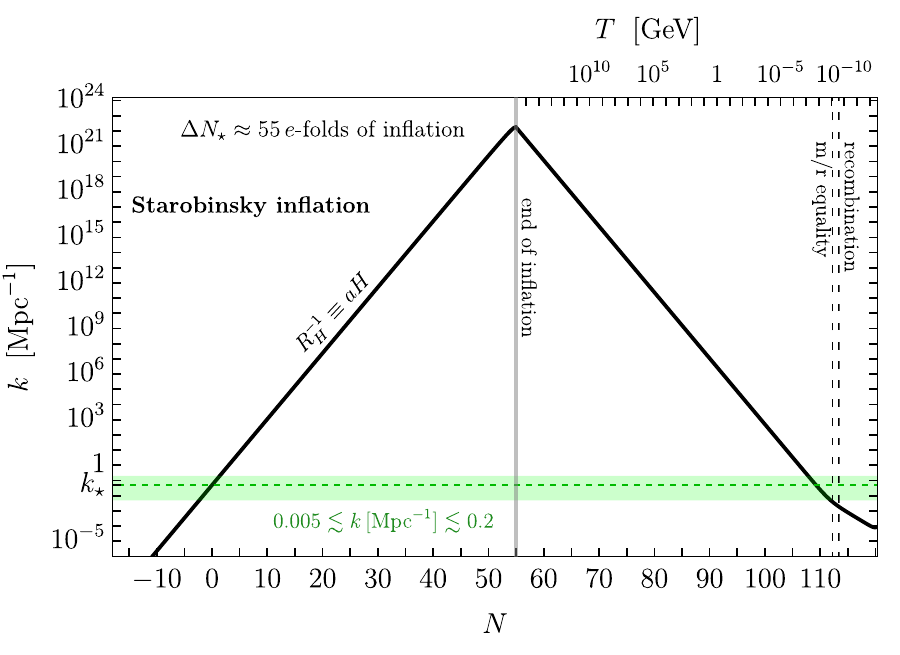}
	\caption{ \it 
 Evolution of the inverse
 comoving Hubble radius. 
 During the inflationary phase, we adopt the Starobinsky model, while the transition to radiation occurs instantaneously. We highlight that CMB scales (green band) re-enter the horizon before recombination epoch.}
\label{fig:StaroPlot}
\end{figure}
To further substantiate this conclusion beyond the slow-roll approximation, we numerically solved the inflationary dynamics by fixing $\Delta N_{\star} = 55$. 
In Fig.\,\ref{fig:StaroPlot} we show the time evolution of the inverse comoving Hubble radius from inflation to the present day. 
The range of comoving wavenumbers that are relevant for CMB observations are indicated with an horizontal green band. 
In particular, $k_{\star} = 0.05$ Mpc$^{-1}$ is indicated with an horizontal green dashed line.
The CMB pivot scale exits the horizon  $\Delta N_{\star} = 55$ $e$-folds before the end of inflation and re-enters the horizon before matter-radiation equality. We assume, consistently with Fig.\,\ref{fig:StaroB}, instantaneous reheating. For completeness, we show in Fig.\,\ref{fig:SolvingHorizonProblem}, how in this model it is possible to solve the horizon problem, simply postulating an additional $\sim 6.75$ \textit{e}-folds of inflation (beyond the 55 that elapse between the exit from the horizon of the CMB scales and the end of inflation).
\section{PBHs production in a toy model}\label{sec:PBHprob}
We now return to the issue of PBH formation from single-field inflation. 
We would now like our inflationary model, in addition to resolving the horizon problem and being compatible with the observed universe, to also produce a sizable population of PBHs.

Working in the scenario in which PBHs form out of the gravitational collapse of large over-densities in the primordial density contrast field\,\cite{Ivanov:1994pa,Ivanov:1997ia}, to achieve a significant amount of dark matter in the form of PBHs, it is necessary for the amplitude of the curvature power spectrum to be around $10^{-2}$ at the relevant range of scales. However, at the scales associated with the Cosmic Microwave Background (CMB), typically around $k\simeq 0.05$ $\textrm{Mpc}^{-1}$, the inflationary power spectrum has an amplitude around $10^{-9}$ \cite{Planck:2018jri}. Therefore, a mechanism is required to enhance the power spectrum at the relevant scales. The above-mentioned enhancement can be dynamically realized in the context of single-field
models of inflation, introducing a phase of ultra slow-roll (USR) during which the inflaton field, after the first conventional phase of slow-roll (SR) that is needed to fit large-scale cosmological observations, almost stops the descent along its potential (typically because of the presence of a quasi-stationary inflection point) before starting rolling down again in a final stage of SR dynamics that eventually ends inflation.

The required USR phase will impose an additional constraint on the model, which, as we will see, essentially reduces the freedom of choice on $\Delta N_{\star}$ that was crucial in the previous example for adapting the inflationary model to observational constraints.
\begin{figure}[!t]
	\centering
\includegraphics[width=0.495\textwidth]{ 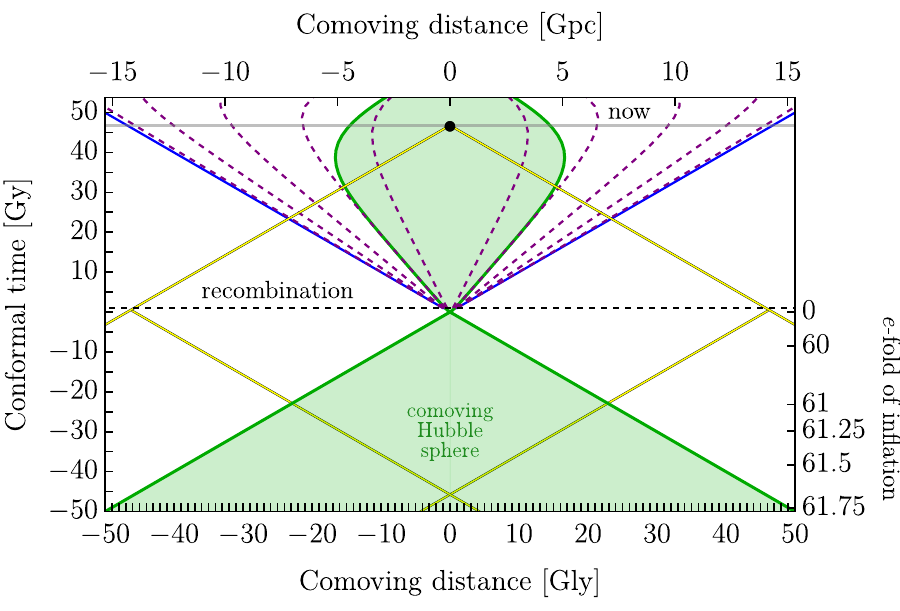}
	\caption{ 
 \it 
 Solution to the horizon problem in the 
 Starobinsky model of inflation, 
 cf. Fig.\,\ref{fig:StaroPlot}.
 }
\label{fig:SolvingHorizonProblem}
\end{figure}

We consider a scalar field theory described, in the absence of gravity, by the following Lagrangian density
\begin{align}
\mathcal{L} = \frac{1}{2}(\partial_{\mu}\phi)(\partial^{\mu}\phi) - V_{\textrm{Staro+dip}}(\phi)\,
\end{align}
where, following Ref.\,\cite{Mishra:2019pzq}, the potential is given by
\begin{align}
V_{\textrm{Staro+dip}}(\phi) \equiv 
V_{\textrm{Staro}}(\phi)\left[
1-A\cosh^{-2}\left(\frac{\phi-\phi_0}{\sigma}\right)
\right]\,.\label{eq:StaroDip}   
\end{align}
The hyperbolic function---characterised by  height $A$, position $\phi_0$ and width $\sigma$ that was added to the Starobinsky model---plays the role of a speed-breaker term for the inflaton.

In the following we introduce some basics notion and equations, useful to determine the power spectrum of the curvature perturbation in presence of single field models with an inflection point.
\subsection{Classical dynamics and power spectra}\label{subsec_MS}

The Hubble-flow parameters $\epsilon_{i}$ (for $i\geqslant 1$) are defined by the recursive relation
\begin{align}
\epsilon_{i} \equiv \frac{\dot{\epsilon}_{i-1}}{H\epsilon_{i-1}}\,,~~~~~\textrm{with:}~~~
\epsilon_0 \equiv \frac{1}{H}\,.\label{eq:HubblePar1}
\end{align}
As customary, we simply indicate as $\epsilon$ the first Hubble parameter, $\epsilon \equiv \epsilon_1 = -\dot{H}/H^2$. 
Instead of the second Hubble parameter $\epsilon_2$, sometimes it is useful to introduce the Hubble parameter $\eta$ 
defined by
\begin{align}
\eta \equiv - \frac{\ddot{H}}{2H\dot{H}} 
= \epsilon - 
\frac{1}{2}\frac{d\log\epsilon}{dN}\,,~~~~
\textrm{with:}~~~\epsilon_2 = 2\epsilon - 2\eta\,.\label{eq:HubblePar2}
\end{align}
Using the number of $e$-folds as time variable,  the inflaton equation of motion reads
\begin{align}
\frac{d^2\phi}{dN^2} + \left[3 - \frac{1}{2\MPl^2}\left(\frac{d\phi}{dN}\right)^2\right]
\left[\frac{d\phi}{dN} + \MPl^2\frac{d\log V(\phi)}{d\phi}
\right] = 0\,,\label{eq:EoM}
\end{align}
and, in turn, the Hubble parameters take the form
\begin{align}
\epsilon  = \frac{1}{2\MPl^2}\left(\frac{d\phi}{dN}\right)^2\,,~~~\eta = 
3 + \frac{\MPl^2(3 - \epsilon)V^{\prime}(\phi)}{V(\phi)(d\phi/dN)}
\,,\label{eq:EoMInfla}
\end{align}
while the Hubble rate is related to the inflaton potential by means of the Friedmann equation
\begin{align}
(3-\epsilon)H^2\MPl^2 = V(\phi)\,.\label{eq:HEvo}
\end{align}
The condition of Ultra-Slow Roll is formally defined as the part of the dynamics for which $\eta > 3/2$\,\cite{Ballesteros:2020qam}.
During this phase, from a classical perspective, the inflaton undergoes a sharp deceleration, nearly coming to a halt. Meanwhile, from a quantum perspective, curvature perturbations experience a phase with a negative friction term in their equation of motion, which significantly boost the power spectrum. This specific feature will be addressed in detail in the following section.

In order to compute the scalar power spectrum of curvature perturbations we need to  solve the Mukhanov-Sasaki (MS) equation \cite{Sasaki:1986hm,Mukhanov:1988jd}
\begin{align}\label{eq:M-S}
\frac{d^2 u_k}{dN^2} &+ (1-\epsilon)\frac{du_k}{dN} + \nn\\
&
\left[
\frac{k^2}{(aH)^2} + (1+\epsilon-\eta)(\eta - 2) - \frac{d}{dN}(\epsilon - \eta)
\right]u_k = 0\,,
\end{align}
with sub-horizon Bunch-Davies initial conditions\,\cite{Bunch:1978yq} at $N \ll N_k$, where $N_k$ indicates the horizon crossing time for the mode $k$, that is the time at which we have $k = a(N_k)H(N_k)$. 
The scalar power spectrum of curvature perturbations  $P_{\zeta}(k)$ is then given by 
\begin{align}\label{eq:PS}
P_{\zeta}(k) = \frac{k^3}{2\pi^2}\left|\frac{u_k(N)}{a(N)\phi^{\prime}(N)}\right|^2_{N > N_{\rm F}(k)}\,
\end{align}
The power spectrum $P_{\zeta}(k)$ does not depend on time because the meaning of Eq.\,(\ref{eq:PS}) is that $P_{\zeta}(k)$ must be evaluated after the time $N_{\rm F}(k)$ at which the mode $|u_k(N)/a(N)\phi^{\prime}(N)|$ freezes to the constant value that is conserved until its horizon re-entry. 
We then define $N_{\rm F}(k) \equiv {\rm max}\{N_k, N_{{\textrm{end}}}\}$, where $N_{{\textrm{end}}}$ marks the end of the USR phase. Modes that exit the horizon before $N_{{\textrm{end}}}$ (i.e., modes with $N_k < N_{{\textrm{end}}}$) are not conserved, even though they are super-horizon, because they subsequently encounter the negative friction phase. Therefore, for these modes, their contribution to Eq.\,(\ref{eq:PS}) must be evaluated at any time $N > N_{{\textrm{end}}} > N_k$ after the USR phase ends. Conversely, modes that exit the horizon after $N_{{\textrm{end}}}$ (i.e., modes with $N_k > N_{{\textrm{end}}}$) freeze to their constant value once they become super-horizon. As a result, the contribution of these 
modes to Eq.\,(\ref{eq:PS}) should be evaluated at any time $N > N_k > N_{{\textrm{end}}}$.

\subsection{PBH abundance and the problem of scales}
When one computes the dark matter fraction consisting in PBHs, the key quantity is the differential mass function describing the fraction of dark matter in the form of PBHs
\begin{equation}
f_{\mathrm{PBH}}(M_{\mathrm{PBH}}) = \frac{1}{\Omega_{\mathrm{CDM}}} \frac{d \Omega_{\mathrm{PBH}}}{d \log M_{\mathrm{PBH}}},
\end{equation}
where $M_{\mathrm{PBH}}$ is the PBH mass,
$\Omega_{\mathrm{CDM}} = \rho_{\mathrm{CDM}}/\rho_{c}$ is  
the ratio of the density of dark matter to the critical density $\rho_c$ today (given by $\Omega_{\mathrm{CDM}} = 0.12\,h^{-2}$, with $h = 0.674$ for the Hubble
parameter) and, similarly, 
$\Omega_{\mathrm{PBH}} = \rho_{\mathrm{PBH}}/\rho_{c}$ is the ratio of the density of matter in the form of PBHs to the critical density today.

A key equation for our discussion is the approximated relation between the horizon mass at time $t$, $M_H(t)$, and the comoving wavenumber $k_H$ of the sourced primordial curvature perturbation that re-enters the horizon at time $t$,
\begin{align}
M_H(t) \sim \left(\frac{10^{6.5}\textrm{Mpc}^{-1}}{k}\right)^2\,M_{\odot}\,,\label{eq:HorizonMasskH}
\end{align}
where $k_H\equiv a(t)H(t)$\,\cite{Kawasaki:2016pql,Sasaki:2018dmp}.
As at the time of formation the mass of PBHs is approximately a fraction of the Hubble horizon mass, Eq.\,(\ref{eq:HorizonMasskH}) relates the mass of PBHs formed from the collapse of curvature perturbations with their comoving wavenumber.

We tune the values of the parameters $(A,\phi_0,\sigma)$ so that the power spectrum of scalar curvature perturbations has an amplitude of the order $10^{-2}$ (cf. the caption of Fig.\,\ref{fig:StaroB}). As we will see explicitly in the next section with the help of an original model, these are the typical values that are needed to form a sizable abundance of PBHs. 
The key observation is that, to accommodate the presence of a sufficiently long USR phase, the value of $\Delta N_{\star}$ is forced to increase. 
This is shown in the same Fig.\,\ref{fig:StaroB}. The red dot corresponds to the case in which the Starobinsky potential is perturbed by the presence of a dip-like feature as in Eq.\,(\ref{eq:StaroDip}).  
The resulting value of $\Delta N_{\star}$ now exceeds the bound given by Eq.\,(\ref{eq:MainLog}).
Note that, on the other hand, in this case the value of $H_{k}$ remains essentially unaltered.
\begin{figure}[!t]
	\centering
\includegraphics[width=0.495\textwidth]{ 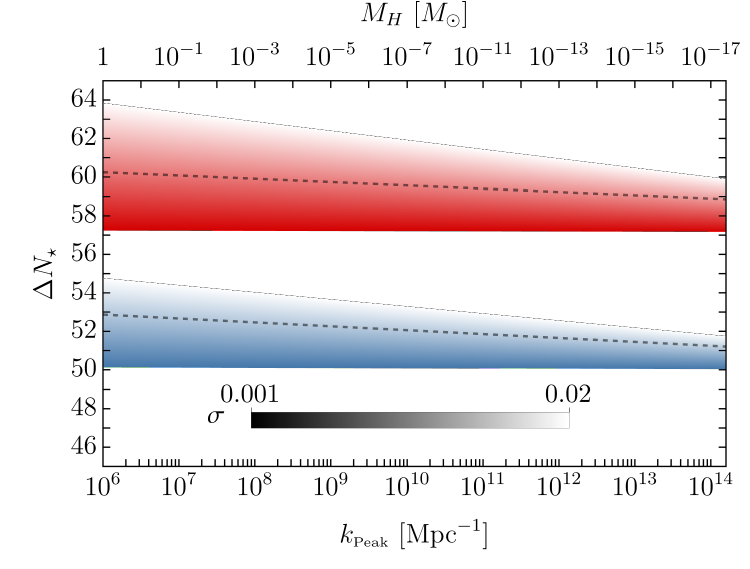}
	\caption{ \it Number of e-folds generated during the inflationary phase by the potential described in Eq.,\ref{eq:StaroDip}, shown for two different sets of inflationary parameters, $(n_s, r)$: $[0.965, 3.5 \times 10^{-3}]$ (red region) and $[0.960, 4.5 \times 10^{-3}]$ (blue region). Given the uncertainties in the computation of PBH abundance (see App.\,\ref{app:PBHabu}), we adopt an "approach-independent" strategy in this plot by fixing the amplitude of the power spectrum at the peak scale, $k_\textrm{peak}$, to $10^{-2}$. The dependence of $\Delta N_\star$ on the model parameter $\sigma$ is shown for the range $\sigma \in [0.001,, 0.2]$. The thicker dashed line corresponds to the case $\sigma = 0.015$, while the thinner dashed line, located at the boundary of the colored regions, corresponds to $\sigma = 0.02$. As $\sigma$ decreases, the number of e-folds rapidly approaches the value obtained in the asymptotic limit of a very narrow peak, i.e., $\sigma \sim 10^{-3}$.
 }
\label{fig:ModelScan}
\end{figure}

\begin{figure}[!t]
	\centering
\includegraphics[width=0.495\textwidth]{ 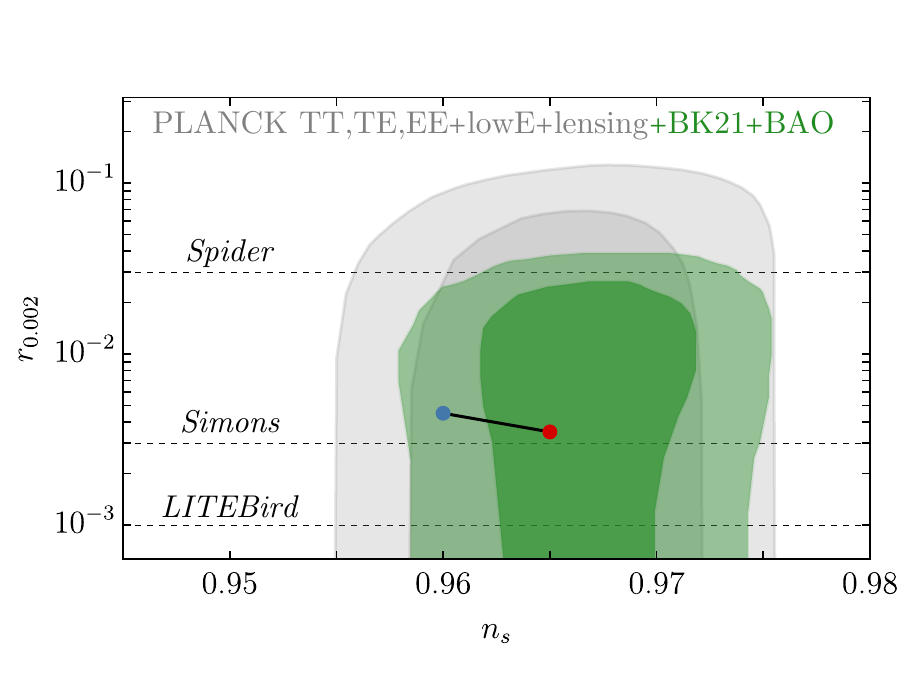}
	\caption{ \it 
Constraints in the $r$ vs. $n_s$ plane using the joint analysis of Planck data (grey region) with additional BICEP/Keck and BAO data (green region) , c.f. \cite{BICEP:2021xfz}. The plot includes 68\% and 95\% confidence contours.
The black line represents the predictions of the toy model discussed in section\,\ref{sec:PBHprob}. Colored dots at the edge of the black line indicate the two regions highlighted in Fig.\,\ref{fig:ModelScan}, following the same color scheme. In addition, we plot the predicted constraints on $r$ from future experiments such as Spider \cite{SPIDER:2017xxz}, Simons Observatory \cite{SimonsObservatory:2018koc}, and LiteBIRD \cite{Matsumura:2013aja}.
 }
\label{fig:ModelsCMBConstraint}
\end{figure}

In Fig.\,\ref{fig:ModelScan}, we show that the issue described in this section persists across a wide range of scales. In particular, we fixed the inflationary parameters $n_s$ and $r$ to be, respectively, $[0.965,3.5*10^{-3}]$ for the red region and $[0.960,4.5*10^{-3}]$ for blue region. Then, we change the position and the shape of the dip, tuning the parameters $\sigma,A$ and $\phi_0$ in Eq.\,\eqref{eq:StaroDip}, in order to have a maximum amplitude of the power spectrum at $10^{-2}$ but at different location, $k_{\rm peak}$. For the first set of parameters (the red region), we observe that $\Delta N_\star$ exceeds the bound given in Eq.\,(\ref{eq:MainLog}) for all the points shown in the plot. However, we note that reducing the parameter $\sigma$ mitigates the increase of $\Delta N_\star$. The parameter $\sigma$ controls the width of the dip in the potential and, consequently, the duration of the USR phase. This confirms that, in this toy model, the excess of e-folds during inflation is primarily determined by the length of the USR phase. Notably, if we instead take as a reference value the second set (blue region), the model becomes compatible with the bound of $\Delta N_\star = 55$ for all values of $\sigma$. Therefore, carefully choosing the parameters of the model (\ref{eq:StaroDip}) 
is possible to evade this issue.
The agreement between the values adopted in Fig.~\ref{fig:ModelScan} and the Planck data\,\cite{Planck:2018jri} is shown in Fig.\,\ref{fig:ModelsCMBConstraint}.

We summarize the main points of this section. For single field inflationary models capable of producing PBHs, it is not enough to verify the compatibility of the amplitude of the power spectrum with the physical constraints and ensure that the PBH production does not exceed the dark matter abundance. Without the requirement of an exotic reheating, i.e $w>1/3$, we need to verify that also the duration of inflation $\Delta N_{\star}$ does not exceed the bound of $\sim55$. As illustrated in Fig.\,\ref{fig:ModelScan}, precise tuning of the model parameters is required to shorten the duration of the USR phase and, ideally, achieve a configuration that avoid the bound mentioned above.

We would like to emphasize that we employ this model solely for the purposes of the above illustrative discussion. We do not consider this phenomenological toy model to realistically describe the effects of an ultra-slow-roll phase in single-field inflation models. For example, it is possible to arbitrarily adjust the position and height of the dip without altering the underlying base potential. 
For this reason, we now move to consider a realistic model of single-field inflation that implements a phase of USR.

\noindent
\section{The double inflection point model}\label{sec:Model}
We present an original model of polynomial single-field inflation capable of producing PBHs across a wide range of scales without requiring a non-minimal coupling with gravity.

Similar to the discussion in the previous section, we consider a scalar field theory described, in the absence of gravity, by the following Lagrangian density
\begin{align}
\mathcal{L} = \frac{1}{2}(\partial_{\mu}\phi)(\partial^{\mu}\phi) - V(\phi)\,,~~~
V(\phi) = \sum_{k=2}^{n}\frac{a_k g^{k-2}}{k! M^{k-4}}\phi^k\,,\label{eq:FlatSpaceTheory}
\end{align}
where $a_k$ are dimensionless numbers, $g$ some  fundamental coupling and $M$ a mass scale.  
We rewrite the potential as
\begin{align}
V(\phi) = \frac{M^4}{g^2} \sum_{k=2}^{n}
\frac{a_k}{k!}\left(\frac{g\phi}{M}\right)^k\,.
\end{align}
It is convenient to rescale the field in units of the mass-to-coupling ratio $M/g$, and we define
$x \equiv g\phi/M$. 
We introduce two approximate stationary inflection points in the above potential. To do this, we find that we need at least $n=6$. We write the potential in the following form
\begin{align}
V(x) = \frac{c_4M^4}{g^2}
\left(
\bar{c}_2 x^2 + 
\bar{c}_3 x^3 +
x^4 +
\bar{c}_5 x^5 + 
\bar{c}_6 x^6
\right)\,,\label{eq:Pot1}
\end{align}
with $c_k \equiv a_k/k!$ and 
$\bar{c}_k \equiv c_k/c_4 = a_k4!/a_4k!$. 
Clearly, the potential in Eq.\,(\ref{eq:Pot1})  renders our theory non-renormalizable and, as such, it should be interpreted as an effective field theory. 
Consequently, we are tacitly operating under the assumption that our inflationary theory is an effective field theory arising from the integration of a sector characterized by a fundamental coupling $g$ and a mass scale $M$. This mass scale corresponds to the typical mass of the heavy degrees of freedom which, in the effective approach, do not participate in the dynamics. 
We might be tempted to impose the condition $g\MPl = M$, which would directly relate the coupling and the mass through the Planck scale. However, for the time being, we will not impose this constraint.

In order to enforce the presence of a double inflection point, we introduce the following parametrization. 
Before proceeding, we remark that this step is not really needed, and one could also work directly with the potential in Eq.\,(\ref{eq:Pot1}). 
However, the parametrization we are about to introduce will be helpful in the course of the numerical analysis.
We impose the four conditions
\begin{align}
V^{\prime}(x_0) =
V^{\prime\prime}(x_0) = 0\,,
~~~
V^{\prime}(x_1) =
V^{\prime\prime}(x_1) = 0\,,\label{eq:ExactStationaryInflectionPoint}
\end{align}
that identify $x_{0,1}$ as exact stationary inflection points. 
Solving for $\bar{c}_k$, we find
\begin{align}
 &V(x) = \frac{c_4 M^4}{g^2} 
\left\{
x^4 + \frac{2}{x_0^2+4x_0x_1 + x_1^2}\times\right. \nn\\
&\left.
\left[
x_0^2x_1^2 x^2 -
\frac{4}{3}x_0x_1(x_0+x_1)x^3 -
\frac{4}{5}(x_0 + x_1)x^5 + \frac{x^6}{3}
\right]
\right\}\,.\label{eq:ExactPotential}
\end{align}
The presence of two inflection points results from a balance between coefficients of opposite signs, specifically those of quadratic/cubic and fifth/sixth order, respectively.

It is known that a shallow local minimum instead of an exact stationary inflection point helps in the production of PBHs, cf. Refs.\,\cite{Ballesteros:2017fsr,Ballesteros:2020qam}. 
For this reason, we opt for a slight generalization of the parametrization given above. 
The simplest possibility is to
perturb the conditions in Eq.\,(\ref{eq:ExactStationaryInflectionPoint}) and write 
$V^{\prime}(x_0) = \kappa\alpha_1$,
$V^{\prime\prime}(x_0) = \kappa\alpha_2$ and 
$V^{\prime}(x_1) = \kappa\alpha_3$, 
$V^{\prime\prime}(x_1) = \kappa\alpha_4$, with $\kappa \equiv c_4M^4/g^2$ (so that $\alpha_i$ are dimensionless numbers).  
Solving again for $\bar{c}_k$, one can then generalize Eq.\,(\ref{eq:ExactPotential}) to the case of approximate stationary inflection points. 
However, the resulting expression is quite cumbersome. For simplicity's sake, therefore, we adopt in our analysis a different parametrization.
We simply write, instead of Eq.\,(\ref{eq:ExactPotential}), the expression 
\begin{align}
V(x) = &\frac{c_4 M^4}{g^2} 
\left\{
x^4 + \frac{2}{x_0^2+4x_0x_1 + x_1^2}\times\right. \nn\\
&\left[
x_0^2x_1^2(1+\beta_2)x^2 -
\frac{4}{3}x_0x_1(x_0+x_1)(1+\beta_3)x^3\right. \nn\\ 
&\left.\left. -
\frac{4}{5}(x_0 + x_1)(1+\beta_5)x^5 + \frac{(1+\beta_6)x^6}{3}
\right]
\right\}\,,\label{eq:DoubleInflectPote}
\end{align}
where the dimensionless coefficients $\beta_i$ parametrize deviations from the situation in which the stationary inflection  points are exact.
For completeness, the mapping between the two representations (that is, between $\alpha_i$ and $\beta_i$) is given in Appendix,\ref{app:Para}.

We minimally couple the theory in Eq.\,(\ref{eq:FlatSpaceTheory}) with Einstein gravity.

We aim to structure our discussion in a broad context, but we will directly present the results of different analyzes along with the corresponding figures focusing on the two realizations of our model. The two cases present the same outcome, i.e. the same PBH mass distribution, but have different durations in e-folds.
In order to construct a discussion around the problem of the mismatch of scales, if not explicitly reported, all the figures from now on are referred to the \textbf{Case A}.
In the first case (\textbf{Case A}) the inflationary lasts too long to have an universe compatible with the usual picture, while in the second case (\textbf{Case B}) the dynamics lasts less than $55$ e-folds, analogously as in the starobinsky case without any dip and there is not any issue related to the standard cosmological evolution. The parameter values are reported in table\,\ref{tab:Parameters}.
\begin{table}[htp]
	\begin{center}
            \textbf{Case A}
		\begin{adjustbox}{max width=.485\textwidth}
		\begin{tabular}{||c||c|c|c|c|c|c|c||}\hline
     \bf{Parameter} &  $x_0$ & $x_1$ & $\beta_2$ & $\beta_3$ & $\beta_5$ & $\beta_6$ & $c_4 g^2$ \\ \hline
      \bf{Value} & 
      1.439
      & 
      6 
      &
      0
      &
      $4.4 \times10^{-4}$
      &
      $-7.05\times10^{-4}$
      &
      $-9.5\times10^{-4}$
      & 
      $4.83\times 10^{-12}$
      \\ \hline\hline
     \bf{Parameter} & $x_0$ & $x_1$ & $\alpha_2$ & $\alpha_3$ & $\alpha_5$ & $\alpha_6$ & $c_4 g^2$ \\ \cline{1-1} \cline{2-8}
      \bf{Value} & 
      1.439
      &
      6
      &
      $-4.3 \times10^{-3}$
      &
      $-3.2\times10^{-3}$ 
      &
      $0.23$
      &
      $0.12$ 
      & $4.83\times 10^{-12}$
      \\ \hline\hline 
     \bf{Parameter} &  &  & $\bar{c}_2$ & $\bar{c}_3$ & $\bar{c}_5$ & $\bar{c}_6$ & $c_4 g^2$ \\ \cline{1-1} \cline{4-8}
      \bf{Value} & 
      &
      & $2.015$
      & $-2.338$
      & $-0.207$
      & $9.4 \times 10^{-3}$
      & $4.83\times 10^{-12}$
      \\ \hline\hline       
      \end{tabular}
	\end{adjustbox}
	\end{center}\vspace{-0.25cm}
       	\begin{center}
        \textbf{Case B}
		\begin{adjustbox}{max width=.485\textwidth}
		\begin{tabular}{||c||c|c|c|c|c|c|c||}\hline
     \bf{Parameter} &  $x_0$ & $x_1$ & $\beta_2$ & $\beta_3$ & $\beta_5$ & $\beta_6$ & $c_4 g^2$ \\ \hline
      \bf{Value} & 
      1.416
      & 
      5.92
      &
      0
      &
      $5.78 \times10^{-4}$
      &
      $-6.0\times10^{-4}$
      &
      $-6.6\times10^{-4}$
      & 
      $4.33\times 10^{-12}$
      \\ \hline\hline
     \bf{Parameter} & $x_0$ & $x_1$ & $\alpha_2$ & $\alpha_3$ & $\alpha_5$ & $\alpha_6$ & $c_4 g^2$ \\ \cline{1-1} \cline{2-8}
      \bf{Value} & 
      1.416
      &
      5.92
      &
      $-3.0 \times10^{-3}$
      &
      $-6.5\times10^{-3}$ 
      &
      $0.20$
      &
      $0.14$ 
      & $4.33\times 10^{-12}$
      \\ \hline\hline 
     \bf{Parameter} &  &  & $\bar{c}_2$ & $\bar{c}_3$ & $\bar{c}_5$ & $\bar{c}_6$ & $c_4 g^2$ \\ \cline{1-1} \cline{4-8}
      \bf{Value} & 
      &
      & $1.991$
      & $-2.325$
      & $-0.208$
      & $9.4 \times 10^{-3}$
      & $4.33\times 10^{-12}$
      \\ \hline\hline       
      \end{tabular}
	\end{adjustbox}
	\end{center}\vspace{-0.25cm}\caption{\it 
Numerical values for the benchmark realizations of our model that is explicitly studied in the course of this work. 
The first set of parameters refers to the potential in Eq.\,(\ref{eq:DoubleInflectPote}).
The second set of parameters refers to the potential discussed in appendix\,\ref{app:Para}. The third set of parameters refers to the potential in Eq.\,(\ref{eq:Pot1}).
    }\label{tab:Parameters}
\end{table}

\subsection{Model Dynamics}
In terms of the a-dimensional field $x \equiv g\phi/M$, Eq.\,(\ref{eq:EoMInfla}) reads 
\begin{align}
\frac{d^2x}{dN^2} + \left[3 - \frac{\xi^2}{2}\left(\frac{dx}{dN}\right)^2\right]
\left[\frac{dx}{dN} + \frac{1}{\xi^2}\frac{d\log V(\phi)}{dx}
\right] = 0\,,
\end{align}
where we introduced the dimensionless ratio $\xi \equiv M/g\MPl$. 
This parameter controls the ratio between the mass scales $M$ and $g\MPl$. Henceforth, we consider the choice $\xi = 1$.
Consequently, we note that the measurement of the amplitude of the power spectrum of curvature perturbations at the CMB scale is sensitive to the combination $c_4g^2$.

\begin{figure}[!t]
	\centering
\includegraphics[width=0.495\textwidth]{ 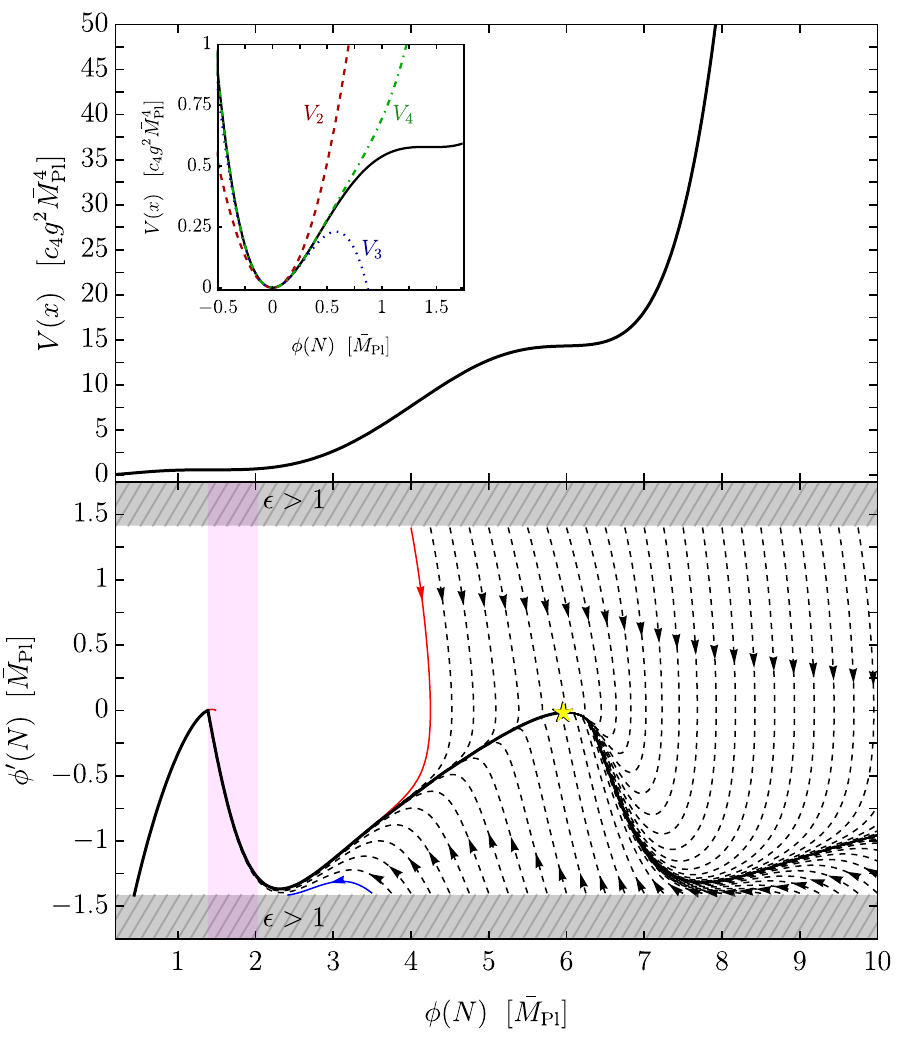}
	\caption{ \it 
  Top panel. 
 Scalar potential given in Eq.\,(\ref{eq:DoubleInflectPote}) with the parameters shown in table\,\ref{tab:Parameters} (\textbf{Case A}). The inset plot shows a zoom-in on the region near the minimum where inflation ends and the reheating phase occurs.
 Bottom panel.
Evolution of the inflationary background from the perspective of its phase space. 
The vertical magenta band indicates the USR phase. 
The yellow star indicates the point where we fit the CMB data.
 }
\label{fig:BackDyn}
\end{figure}

We show the background dynamics in Fig.\,\ref{fig:BackDyn}.
The crucial point of this part of the analysis is to determine whether the first approximate inflection point results in a non-attractive phase of inflation. 
To explore this aspect, we examine in Fig.\,\ref{fig:BackDyn} the phase space of the inflationary dynamics. 
The upper part of the figure depicts the potential and clearly illustrates the presence of the two stationary approximate inflection points. 
In the lower part of the figure, we present the corresponding phase space obtained by numerically solving the equation of motion for the inflaton for generic initial positions and velocities. 
First, we consider the dynamics close to the first quasi-inflection point (located at $x_1 = 6$). 
The inflationary trajectory remains an attractor despite the inflaton experiencing a sharp deceleration when passing through the approximate inflection point. 
Thereafter, the inflaton begins to accelerate again until it encounters the second approximate stationary inflection point, where---like in a cosmic roller coaster---it experiences a second, and even sharper, deceleration. 
As we shall see, this part of the dynamics is crucial for the formation of PBHs. 
As well known, this second part of the dynamics features a breaking of the attractor phase\,\cite{Morse:2018kda,Cai:2018dkf,Dimopoulos:2017ged,Biagetti:2018pjj} (cf. also Ref.\,\cite{Pattison:2018bct} for a critical assessment of the statement).
In our discussion, we observe that our phase space portrait is bounded, at the upper part of the diagram relative to the attractor trajectory, by the red-marked trajectory, where the inflaton remains trapped in the minimum formed by the second approximate stationary inflection point.
Similarly, at the lower part, along the trajectory marked in blue, inflation ends ($\epsilon > 1$) before the inflaton reaches the absolute minimum of the potential. 
These limitations, as mentioned, are typical of the presence of a nearly stationary inflection point required to produce a sizable abundance of PBHs. However, the point that we wish to emphasize is that the presence of the first nearly stationary inflection point, which corresponds to reproducing the CMB, does not alter the attractor dynamics of the inflationary solution.

In Fig.\,\ref{fig:BackDyn}, the yellow star corresponds to the point in phase space where we reproduce the CMB data in the \textbf{Case A} of our model, cf. table\,\ref{tab:Parameters}.

In the solution presented in Fig.\,\ref{fig:BackDyn}, which is relevant for the formation of sub-solar mass PBHs, we highlight the presence of an USR phase through a vertical magenta band. 

\begin{figure*}[!t]
	\centering
\includegraphics[width=0.995\textwidth]{ 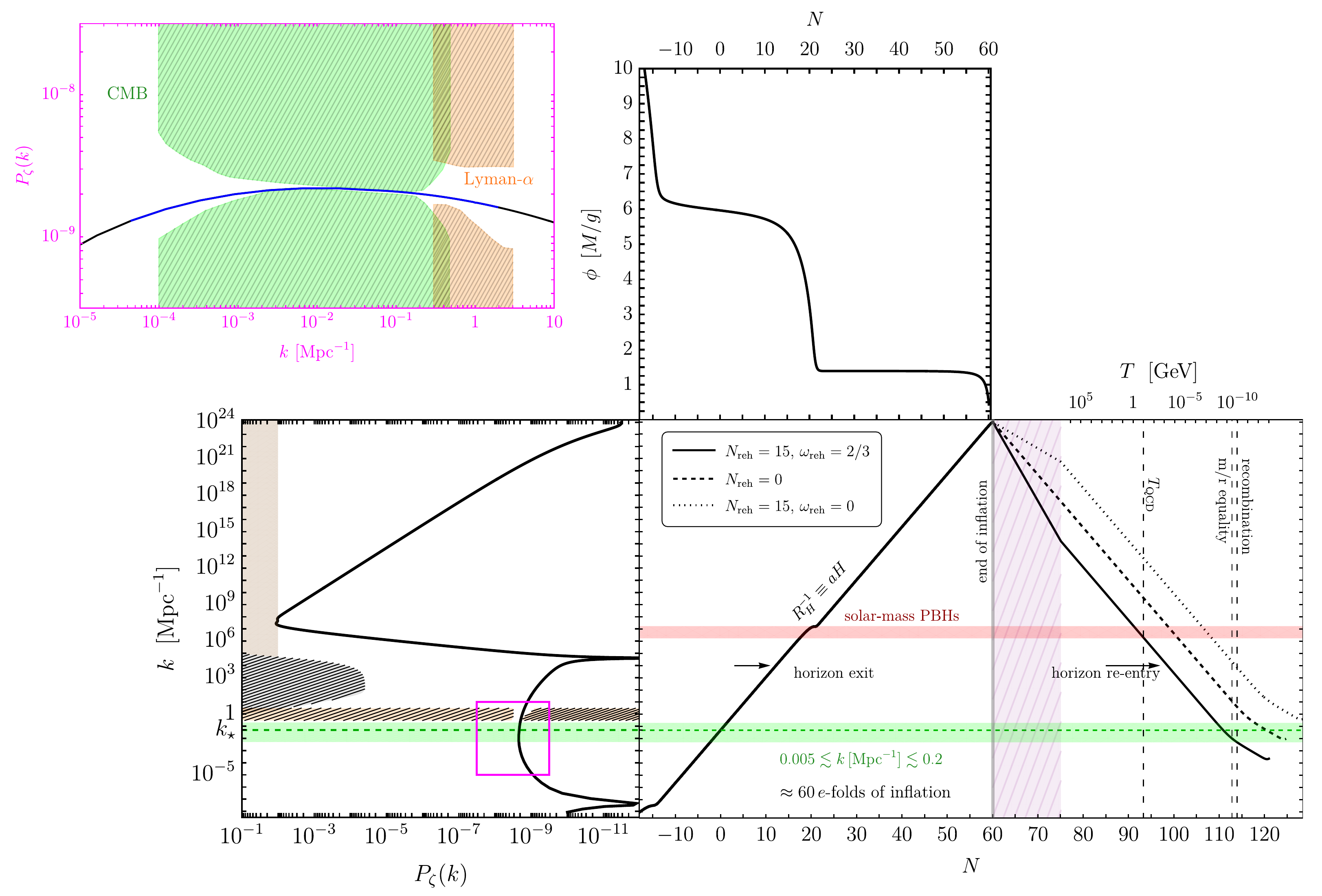}
	\caption{\it Analysis performed for the \textbf{case A} in Tab.\,\ref{tab:Parameters}.
Top left panel (with magenta borders). Power spectrum of curvature perturbations as a function of comoving wavenumbers, zoomed in on the region of scales relevant for CMB measurements. 
$\perp$-shaped central panel.
On the left, we show the power spectrum of curvature perturbations over the entire range of scales covered by the inflationary dynamics.  
We plot the region excluded by CMB anisotropy measurements, Ref.\,\cite{Planck:2018jri}, the FIRAS bound on CMB spectral distortions, Ref.\,\cite{Chluba:2012we} (see also Ref.\,\cite{Jeong:2014gna,Iovino:2024tyg}) and the bound obtained from Lyman-$\alpha$ forest data \cite{Bird:2010mp}.  
At the top, we show the solution of the background equation during the inflationary phase. 
In the center-right, we show the evolution of the inverse comoving Hubble radius during inflation (center) and in the subsequent phase (right). 
In the post-inflationary phase, we show different realizations of reheating (see legend for details). The temperature evolution (top $x$-axis) refers to the solution with 
$\omega_{\textrm{reh}} = 2/3$ and 
$N_{\textrm{reh}} = 15$. 
Similarly, the three vertical lines---indicating, from left to right, the QCD phase transition, matter-radiation equality, and recombination---refer to the solution with $\omega_{\textrm{reh}} = 2/3$ and 
$N_{\textrm{reh}} = 15$.
 }
\label{fig:FullUniEvo}
\end{figure*}

Now we apply the formalism of the Mukhanov-Sasaki equation (see Sec.~\ref{subsec_MS}) in order to compute the scalar power spectrum to the model discussed in this section. In the upper-left panel of Fig.\,\ref{fig:FullUniEvo} (with magenta-colored edges), we show one of the key result of our analysis. 
What we present is the power spectrum of curvature perturbations within the comoving wavenumber window 
$10^{-4} \lesssim 
k\,\,[\textrm{Mpc}^{-1}] \lesssim 1$. 
This range is of outmost importance for validating the model, as it is highly constrained by measurements from the CMB and the Lyman-$\alpha$ forest. 
We fit the power spectrum of the model (considering its realization given in table\,\ref{tab:Parameters}) with the functional form
\begin{align}\label{eq:ParametricPS}
P_{\zeta}(k) =
A_s\left(\frac{k}{k_{\star}}\right)^{
{n_s - 1 +\frac{\alpha}{2}\log\frac{k}{k_*}+
\frac{\vartheta}{6}\log^2\frac{k}{k_*}+\dots}
}\,,
\end{align}      
with $\alpha = dn_s/d\log k$, $\vartheta= d^2n_s/d\log k^2$.
Generally, these inflationary parameters are evaluated on the pivot scale $k_* = 0.05$ Mpc$^{-1}$, while the tensor-to-scalar ratio, which is relevant for the comparison with BK18, is calculated at the reference scale $k=0.002$ Mpc$^{-1}$. Using the slow-roll approximation, the expression is given by
\begin{align}
r_{0.002}  = 16\,\epsilon(N_{0.002})\,,\label{eq:Tensor2ScalarValue}
\end{align}
where $N_{0.002}$ is the \textit{e}-fold time at which the reference scale $k=0.002$ $\textrm{Mpc}^{-1}$ crosses the comoving Hubble horizon.

The values of the inflationary parameters for both cases are reported in Tab.~\ref{tab:InfParameters}.
\begin{table}[htp]
	\begin{center}
		\begin{adjustbox}{max width=.485\textwidth}
		\begin{tabular}{||c||c|c|c|c|c|c||}\hline
     \bf{Parameter} &  $A_s$ & $n_s$ & $\alpha$ & $\theta$ & $r$ & $N$ \\ \hline
      \bf{Case A} & 
      $2.15\times 10^{-9}$
      & 
      0.965 
      &
      -0.0288
      &
      0.0016
      &
      $2.20 \times 10^{-3}$
      &
      60
      \\ \hline
      \bf{Case B} 
      & 
      $2.15\times 10^{-9}$
      & 
      0.959
      &
      -0.0275
      &
      0.0013
      &
      $1.87 \times 10^{-3}$
      &
      53.5
      \\ \hline\hline       
      \end{tabular}
	\end{adjustbox}
	\end{center}\vspace{-0.25cm}\caption{\it 
Inflationary parameters for the two benchmark cases under considerations reported in Tab.\,\ref{tab:Parameters}.
    }\label{tab:InfParameters}
\end{table}

As expected, the model is characterized by a sizable running of the spectral index.
For comparison, we quote (at the 68\% confidence level) the value of the running parameters $\alpha = 0.0011 \pm 0.0099\,, 
\ \vartheta = 0.009 \pm 0.012\ $\,\cite{Planck:2018vyg}.
Therefore, our model remains within the 3-$\sigma$ confidence interval.

Regarding the value of $r$, it is in perfect agreement with the limits of 
BK18 ($r \lesssim  0.036$ at 95\%). This is one of the advantages of having an almost stationary inflection point at CMB scales, as in this case $\epsilon$ becomes very small, thereby dragging along the value of $r$.

\begin{figure}[!t]
	\centering
\includegraphics[width=0.495\textwidth]{ 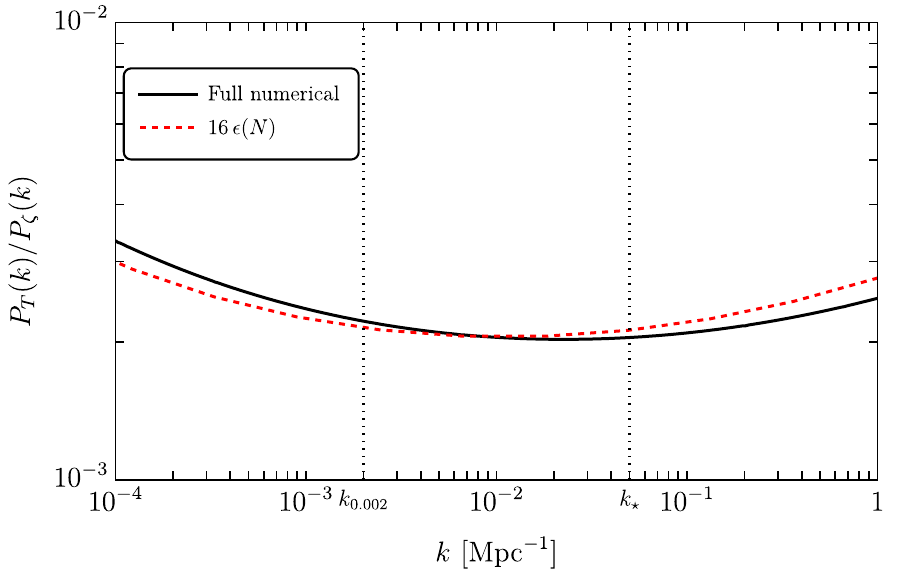}
	\caption{ \it 
 Tensor-to-scalar ratio $P_T(k)/P_{\zeta}(k)$ compared to the consistency relation 
$r=16\epsilon$.
 }
\label{fig:Tensor2Scalar}
\end{figure}

Beyond the slow-roll approximation, the power spectrum of tensor perturbations $P_T(k)$ is given by 
\begin{align}
P_T(k) = 
\frac{4k^3}{\pi^2}\left|
\frac{v_k(N)}{a(N)\MPl}
\right|^2_{N > N_k}\,,
\end{align}
where the mode functions $v_k(N)$ solve the differential equation
\begin{align}
\frac{d^2v_k}{dN^2}
+ 
(1-\epsilon)\frac{dv_k}{dN}
+
\left[
\frac{k^2}{(aH)^2}
- (2-\epsilon)
\right]v_k = 0\,,
\end{align}
with sub-horizon Bunch-Davies initial conditions.
In Fig.\,\ref{fig:Tensor2Scalar} we show the scale dependence of the tensor-to-scalar ratio $P_T(k)/P_{\zeta}(k)$ compared to the consistency relation 
$r=16\epsilon$.
The numerical analysis confirms that the estimate of $r$ given in Eq.\,(\ref{eq:Tensor2ScalarValue}) is an excellent approximation of the full numerical solution.

Let us now extend our analysis to shorter scales.
In Fig.\,\ref{fig:FullUniEvo}, in the left corner of the $\perp$-shaped central panel, we show the full power spectrum of curvature perturbations. 
The part of the power spectrum corresponding to the short scales of the CMB, analyzed previously, is highlighted with a magenta box. 
As mentioned earlier, the dynamics is tuned to produce a sizable enhancement of the power spectrum at scales 
$k = O(10^6\textrm{-}10^8)$ Mpc$^{-1}$.

\subsection{(Sub)solar-mass primordial black holes}\label{sec:SubSolarPBH}
For the computation of the PBHs mass fraction $f_{\mathrm{PBH}}(M_{\mathrm{PBH}})$ we follow the threshold statistics formalism\,\cite{Young:2019yug,DeLuca:2019qsy} (see Appendix\,\ref{app:PBHabu} for technical details).

\begin{figure}[!t]
	\centering
\includegraphics[width=0.495\textwidth]{ 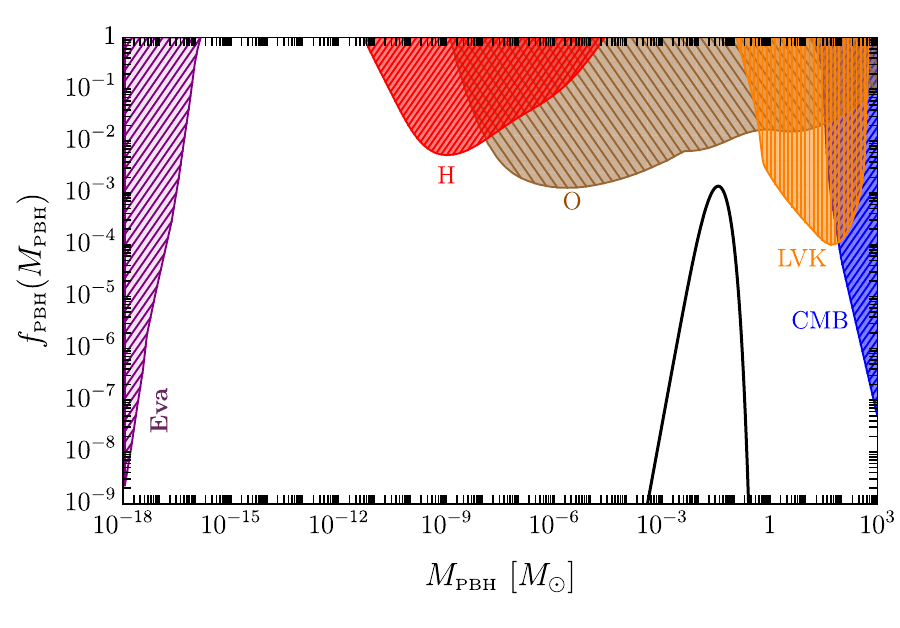}
	\caption{ \it 
Fraction of dark matter in the form of PBHs with mass $M_{\textrm{PBH}}$. We show the following most constraining bounds:
\textbf{Ev}aporation constraints (see also \cite{Saha:2021pqf,Laha:2019ssq,Ray:2021mxu}): EDGES\,\cite{Mittal:2021egv}, 
CMB\,\cite{Clark:2016nst}, INTEGRAL\,\cite{Laha:2020ivk,Berteaud:2022tws}, 511 keV\,\cite{DeRocco:2019fjq,Dasgupta:2019cae}, Voyager\,\cite{Boudaud:2018hqb}, 
EGRB\,\cite{Carr:2009jm});
microlensing constraints from the Hyper-Supreme Cam (\textbf{HSC}), Ref.\,\cite{Niikura:2017zjd}; 
microlensing constraints from \textbf{O}GLE, Refs.\,\cite{Niikura:2019kqi,Mroz:2024wag,Mroz:2024wia};  
constraints from modification of the \textbf{CMB} spectrum due to accreting PBHs, Ref.\,\cite{Serpico:2020ehh,Agius:2024ecw};
direct constraints on PBH-PBH mergers with \textbf{L}IGO, Refs.\,\cite{Andres-Carcasona:2024wqk} (see also \cite{LIGOScientific:2019kan,Kavanagh:2018ggo,Wong:2020yig,Hutsi:2020sol,DeLuca:2021wjr,Franciolini:2021tla,Franciolini:2022tfm}).}  
\label{fig:PBHAbu}
\end{figure}

We present our findings in Fig.\,\ref{fig:PBHAbu} where the PBH mass function peaks around the (sub)solar mass region. 

These PBHs are potentially highly interesting from a phenomenological perspective. On the one hand, considering a binary system of PBHs, a merger event could be detectable by the gravitational interferometers of the LIGO/Virgo/KAGRA collaboration; up to now, specifically targeted searches of subsolar mass compact objects, which would provide a smoking gun signal of the existence of PBHs, have been unsuccessful~\cite{LIGOScientific:2018glc, LIGOScientific:2019kan, Nitz:2020bdb, Nitz:2021vqh, Nitz:2022ltl, Miller:2021knj, Morras:2023jvb, Mukherjee:2021ags,Phukon:2021cus, Mukherjee:2021itf, Andres-Carcasona:2022prl, Miller:2024fpo,Crescimbeni:2025ywm}\footnote{Candidate sub-solar events have been claimed in the literature~\cite{Prunier:2023uoo, Morras:2023jvb}, although without sufficient statistical evidence.}. On the other hand, the same perturbations that would generate the PBHs could serve as a source for a stochastic background of gravitational waves at the nHz frequency, potentially consistent with that recently observed by PTA experiments\,\cite{Franciolini:2023pbf}. 

In this mass range, experimental constraints allow for at most an order of $10^{-3}$ fraction of dark matter in the form of PBHs\footnote{For simplicity constraints obtained for monochromatic mass function are shown. Since the mass function is quite narrow, we expect that the modification of these constraints is negligible\,\cite{Carr:2017jsz,Andres-Carcasona:2024wqk}.}.

On its own, this could be an interesting result, because it is not straightforward to construct single-field inflation models that can produce a sizable abundance of solar-mass PBHs without conflicting with predictions at CMB scales (considering solely inflationary parameters such as the spectral index or tensor-to-scalar ratio).

\subsection{A mismatch of scales}

The cosmological evolution of the comoving Hubble radius plotted in Fig.\,\ref{fig:FullUniEvo} clearly illustrates the main point of our analysis.
In the \textbf{Case A} to accommodate the USR dynamics necessary for PBH formation the inflationary period, extending from the moment the CMB scales exit the horizon to the end of inflation itself, is constrained to last approximately 60 e-folds. This implies that the CMB scales are stretched to the point where they do not have sufficient time to re-enter the horizon consistently with the post-inflationary evolution of the universe. 
This is evident when considering the evolution of the comoving Hubble radius in relation to the comoving scales of the CMB (shown as a green band) in the plot of Fig.\,\ref{fig:FullUniEvo}. Specifically, we first consider the case where the post-inflationary evolution of the universe is characterized by an instantaneous reheating process. In this case, the evolution of the Hubble radius after inflation is governed by the radiation content of the universe and is shown as a dashed black line. 
To draw this line, we start from the value of $aH$ at the end of inflation. 
The entire evolution of $H$ is given by Eq.\,(\ref{eq:HEvo}) once the overall scale of the potential, set by $c_4g^2$ in units of $\MPl^4$, is fixed by imposing the correct normalization of the curvature power spectrum at the CMB pivot scale. In terms of the value of the scale factor, we find $a_{\textrm{end}} = k_{\star}e^{N_{\textrm{end}}-N_{\star}}/H_{\star}$.
Under the assumption of instantaneous reheating, we can use Eq.\,(\ref{eq:H0EvoStandard}) to follow its subsequent temporal evolution. 
The latter is completely nailed down by standard cosmology. 
Specifically, starting from $H$ at the end of inflation and arriving at the observed value $H_0$ today completely determines the number of $e$-folds of post-inflationary evolution. We remark that in this part of the analysis we do {\it not} normalize the scale factor in such a way that $a_0 = 1$. In contrast, the normalization of the scale factor is carried out implicitly considering the match between the standard cosmological evolution in Eq.\,(\ref{eq:H0EvoStandard}) and the value of $aH$ at the end of inflation, as discussed above.

We observe how we arrive at a universe at the time of recombination in which the scales of the CMB are still outside the horizon at the time of matter-radiation equality. Therefore, the dashed black line is clearly incompatible with the cosmology of our universe. 

If we consider a reheating scenario described by a perfect fluid phase with $\omega_{\textrm{reh}} < 1/3$ (dotted line in the post-inflationary evolution of the comoving Hubble radius in Fig.\,\ref{fig:FullUniEvo}), the situation worsens.
To clarify, let us consider the case with $\omega_{\textrm{reh}} = 0$, corresponding to the situation where the reheating phase is matter dominated. 
In a matter-dominated reheating phase, the comoving Hubble radius expands slower compared to radiation domination, exacerbating the tension with standard cosmological evolution. 
Indeed, if we attempt to connect the value of the Hubble rate $H$ at the end of the matter-dominated reheating phase with the observed value today, we find ourselves again in a situation where the scales of the CMB are well outside the horizon at the time of matter-radiation equality.

The only way out seems to be to postulate a reheating phase where the universe's dynamics are dominated by a perfect fluid with $\omega_{\textrm{reh}} > 1/3$. In this case, it is possible to counterbalance the effect of the USR phase---since now during reheating the comoving Hubble radius expands faster compared to radiation domination---and connect, at the end of reheating, to a universe consistent with standard cosmology.

This situation corresponds to the solid black line in the post-inflationary evolution of the comoving Hubble radius in Fig.\,\ref{fig:FullUniEvo}. 
As seen, in this case the scales of the CMB manage to re-enter the horizon just before matter-radiation equality. 
Note how, in the example shown, 15 $e$-folds of reheating with $\omega_{\textrm{reh}} = 2/3$ are 
required. 
It is natural to wonder how realistic this requirement is. 
A reheating phase with $\omega_{\textrm{reh}}>1/3$ is a somewhat exotic possibility because, as we have previously discusses, it would require 
a scalar potential dominated in the minimum by non-renormalizable operators.

As previously discussed, \textbf{Case B} features a configuration of potential parameters that results in the same mass distribution as \textbf{Case A} but fewer than 55 e-folds of inflation, thereby avoiding issues related to standard cosmological evolution. Unlike the Starobinsky+dip model, the relationship between the model parameters and the duration of the inflationary stage, $\Delta N_\star$, is non-trivial, as shown in Tab.\,\ref{tab:Parameters}. Consequently, it is not possible to produce a plot analogous to Fig.\,\ref{fig:ModelScan}.
For polynomial single-field models, such as the one presented in this section, assuming equal configurations at the CMB scales, the total number of e-folds of inflation is related to the value of the inflationary parameter $\eta$ during the USR phase ($\eta_{\rm USR}$), its duration ($\Delta N_{\rm USR}$) and the subsequent slow-roll phase ($\eta_{\rm SR}$). By carefully tuning the model parameters, it is possible to obtain, with a different set of values of ($\eta_{\rm USR},\eta_{\rm SR},\Delta N_{\rm USR}$), the same PBH  mass function while simultaneously reducing the total number of e-folds. Consequently, even for polynomial single-field inflationary models, we can evade the bound given by Eq.\,\ref{eq:MainLog} without requiring an exotic reheating phase, i.e. $\omega_\textrm{reh} > 1/3$.

In the following, we present a more concrete discussion on reheating, where we examine a more explicit realization of its dynamics.

\subsection{Modeling the dynamics of reheating}\label{sec:DynaReh}
To strengthen our previous discussion, we go beyond the approximation in which reheating is described by a phase dominated by a perfect fluid with a fixed $\omega_{\textrm{reh}}$. 
Naturally, this requires introducing some model dependence. Specifically, we introduce a reheating portal by Yukawa coupling the inflaton to a fermion. 
At the tree level, the inflaton decay width reads $\Gamma_{\phi} = y^2m_{\phi}/8\pi$ with Yukawa coupling $y$. In our model, the inflaton mass takes the value
\begin{align}
m_{\phi}^2 = \frac{4c_4 g^2(1+c_2)x_0^2 x_1^2\MPl^2}{x_0^2 + 4x_0x_1 +x_1^2}\,.
\end{align}
The dynamics of the reheating stage is described by the system
\begin{align}
\dot{\rho}_{\gamma} + 4H\rho_{\gamma} - \Gamma_{\phi}\dot{\phi}^2 & = 0\,,\\
\ddot{\phi} + 3H\dot{\phi} + 
\Gamma_{\phi}\dot{\phi} + \frac{dV}{d\phi} & = 0\,,\\
3\MPl^2 H^2 & =  
\left[
\frac{\dot{\phi}^2}{2} + V(\phi) + \rho_{\gamma}
\right]\,,\label{eq:FridReah}
\end{align}
that can be easily solved by imposing as initial conditions the value of $\phi$ and $\dot{\phi}$ at the end of the inflationary stage (with zero initial radiation energy density). 
The damping factor $\Gamma_{\phi}$ takes into account the decay of the inflaton and represents the transfer of energy from the inflaton field to radiation. 
We keep track of $\omega_{\textrm{reh}}$ by writing
\begin{align}
\omega_{\textrm{reh}}(N) = 
\frac{
\dot{\phi}^2/2 - V(\phi) + \rho_{\gamma}/3
}{
\dot{\phi}^2/2 + V(\phi) + \rho_{\gamma}
}\,.\label{eq:OmegaRehEvo}
\end{align}
\begin{figure}[!t]
	\centering
\includegraphics[width=0.495\textwidth]{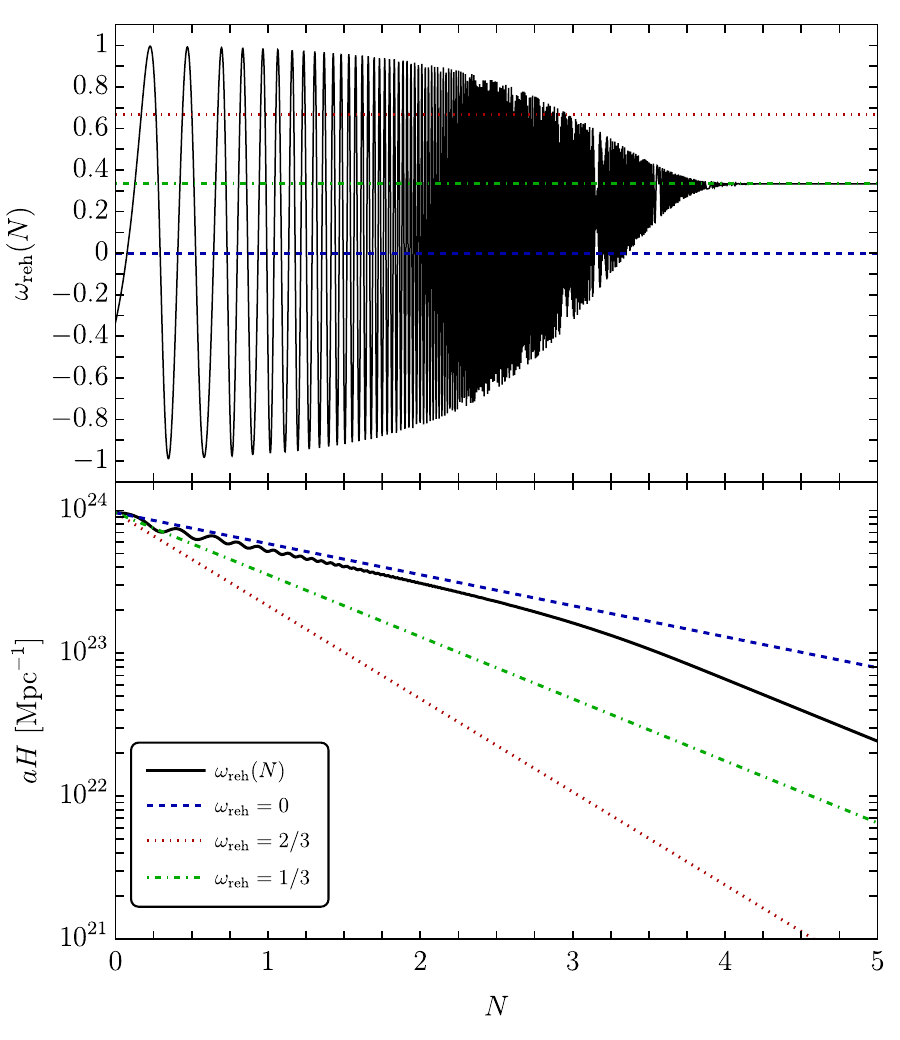}
	\caption{ \it 
 Top panel. 
 Time evolution of the equation of state parameter during reheating, cf. Eq.\,(\ref{eq:OmegaRehEvo}). 
 We take $y=0.25$, and $N=0$ corresponds to the end of inflation. 
 Bottom panel.
 Time evolution of the inverse comoving Hubble radius. We consider 
 $\omega_{\textrm{reh}} = 0,1/3,2/3$ and 
 the case discussed in section\,\ref{sec:DynaReh}, in which we model the dynamics of reheating.
 }
\label{fig:OmegaEvo}
\end{figure}
The result is shown in Fig.\,\ref{fig:OmegaEvo}. 
In the top panel, we show the evolution of $\omega_{\textrm{reh}}$ (with $N=0$ the beginning of the reheating stage).
$\omega_{\textrm{reh}}$ starts from 
$-1/3$ (the value at the end of inflation) and rapidly oscillates in time until the energy stored in the inflaton field gets entirely converted into radiation ($\omega_{\textrm{reh}} = 1/3$).
Note how during the reheating phase, $\omega_{\textrm{reh}}$ can instantaneously be greater than $1/3$ (and saturate the causality limit $\omega_{\textrm{reh}} = 1$). 
In the bottom panel, we show the evolution of the (inverse) comoving Hubble radius computed according to Eq.\,(\ref{eq:FridReah}).

The evolution of $aH$ resembles that of a matter-dominated epoch during reheating, before decaying to radiation. This is expected, as the inflaton transfers its energy to radiation through oscillations around a minimum dominated by the quadratic term of the potential.

This example clearly demonstrates that persisting in the use of a single-field model aiming to unify both inflationary and reheating dynamics makes it extremely challenging to resolve the scale mismatch described in the preceding section. 
\section{Conclusions}\label{sec:Conc}
The study of primordial black holes formed from the collapse of cosmological perturbations requires that the power spectrum of the scalar curvature perturbation field aligns with observational constraints across various scales, such as those provided by CMB observations\,\cite{Planck:2018jri}, FIRAS\,\cite{Chluba:2012we} and  Lyman-$\alpha$ forest\,\cite{Bird:2010mp}. Furthermore, it is crucial to avoid overproducing PBHs relative to the dark matter abundance observed today and to prevent excessive scalar-induced gravitational waves that would conflict with non-detections from experiments like the Pulsar Timing Array collaborations,\cite{Iovino:2024tyg}.

In this work, we explored the compatibility of single-field inflationary models, featuring an ultra-slow-roll phase, with standard cosmology. Specifically, we imposed the condition that, at the time of recombination, the size of the comoving scales should be smaller than those relevant for CMB observations ($k\sim[0.005,0.02]\,\textrm{Mpc}^{-1}$). This translates into an upper bound on the duration of the inflationary phase, formulated in a model-independent way through Eqn.~\eqref{eq:MainLog}. Assuming scenarios of instantaneous or standard reheating (i.e., $-1/3 \leqslant \omega_{\textrm{reh}} \leqslant 1/3$), Eqn.~\eqref{eq:MainLog} limits the total number of \textit{e}-folds after the horizon exit of modes with inverse wavelength $k\approx0.05\,\textrm{Mpc}^{-1}$ to be at most 55, but more than $50$ in order to solve the horizon problem. Once a specific inflationary model is chosen, Eqn.~\eqref{eq:MainLog} translates into a constraint on the model parameter space. We first illustrate our criterion using a simple toy model: Starobinsky inflation with a dip feature. We then introduce a more realistic scenario involving two inflection points in the scalar potential. The latter model is particularly interesting as constructing single-field inflation capable of generating a sizable population of solar-mass PBHs without conflicting with CMB-scale constraints is nontrivial. Using both models, we demonstrated that, fixing the values of the inflationary outcomes ($n_s$, $r$) compatible with Planck+BICEP/Keck data and a similar PBH mass distribution, the constraint on the total duration of inflation could exclude certain parameter configurations. For istance, in Fig.~\ref{fig:FullUniEvo} we showed that configuration \textbf{A} is not viable for standard reheating scenarios, although compatible with CMB observations.
We stress that the constraint for a given model can not always be
visualized simply as in Fig.~\ref{fig:ModelScan}. This is because the actual number of e-folds is intrinsically tied to the specific form of the potential and is thus strongly dependent on the parameters of the model. In particular, for the double-inflection-point model, identifying viable configurations fully consistent with Planck constraints and capable of producing (sub-)solar-mass PBHs proves technically challenging.

One way to relax our constraint is by assuming an exotic reheating scenario with $\omega_{\textrm{reh}} > 1/3$. During reheating, the equation of state parameter can transiently exceed $1/3$ due to various mechanisms. In perturbative reheating, this condition arises if the potential near its minimum is dominated by higher-dimensional operators, $V(\phi) \propto \phi^{p}$ with $p>4$, a scenario that, however, conflicts with naturalness. 
A phase of exotic reheating is also realized in the context of non-oscillatory or quintessential inflation models\,\cite{Peebles:1998qn,Dimopoulos:2001ix}.
This class of models is extremely interesting because it allows for a unified description of both the phase of inflationary expansion and the phase of late dark energy. 
In general terms, in these models, the inflaton potential undergoes a sudden drop, which is responsible for the fast-roll of the scalar field after the end of inflation. On both sides of this drop, the inflaton potential presents two asymptotically flat regions: the inflationary plateau and the quintessence tail. 
In the quintessence tail, the dynamics of the scalar field are dominated by the kinetic term, leading to a phase of kination dominance. 
Therefore, one might consider the production of PBHs within the context of these models. 
In this regard, a potential difficulty could be associated with the necessity of having trans-Planckian field excursions. 
During a phase of kination dominance, in fact, the Friedmann equation in terms of the Hubble parameter 
$6\MPl^2H^2 = \dot{\phi}^2$ implies 
\begin{align}
\Delta\phi \approx \sqrt{6}\MPl \Delta N\,,
\end{align}
meaning that a canonically normalized scalar field varies by $O(\MPl)$ during each $e$-fold of kination. 
Since in our case the inflaton completes the inflationary dynamics at sub-Planckian values, connecting our potential to a quintessence tail does not seem straightforward. 
Undoubtedly, this path could be relevant to explore in future work. 

Whether the exotic reheating is constituted by a kination dominance or other scenarios (see for example gravitational reheating\,\cite{Ford:1986sy} or oscillon-mediated reheating\,\cite{Amin:2014eta,Lozanov:2017hjm}), in any case, one needs to separate the inflationary dynamics responsible for boosting the power spectrum from those governing reheating, treating the latter in an ad hoc manner to resolve the scale issue.

Another possibility involves reheating mediated by spectator fields\,\cite{Bassett:1997az,Tsujikawa:1999jh}, coupled to the inflaton, which subsequently transfer energy to Standard Model particles. However, this requires introducing new particles beyond the Standard Model.

\acknowledgments
\noindent
We thank G. Franciolini, A. Riotto and M. Valli for useful discussions.
This work is partially supported by
ICSC - Centro Nazionale di Ricerca in High Performance Computing, Big Data and Quantum Computing, funded by European Union-NextGenerationEU.
A.J.I. thanks the University of Geneve for the nice hospitality during the realization of this project. 
L.D.G. is supported by NSF Grants No. PHY-
2207502, AST-2307146, PHY-090003 and PHY-20043, by
NASA Grant No. 21-ATP21-0010, by the John Templeton
Foundation Grant 62840, and by the Simons Investigator
Grant No. 144924. 
\newpage
\setcounter{equation}{0}
\setcounter{section}{0}
\setcounter{table}{0}
\setcounter{figure}{0}
\makeatletter
\appendix
\section{A tale of scales}\label{app:Tales}
In this appendix, we present the derivations of the expressions used in section II.

The initial step of our discussion pertains to the evolution of the post-inflationary universe. In this case, we adopt the standard $\Lambda$CDM cosmology. 
We use the numerical values of the astrophysical constants and parameters reviewed in Ref.\,\cite{Workman:2022ynf} (cf. also 
\href{https://pdg.lbl.gov/2023/reviews/contents_sports.html}{pdg tables}).

The evolution of the Hubble expansion rate in $\Lambda$CDM cosmology follows from the Friedmann equation, and takes the form  
\begin{align}
H(a) = H_0\sqrt{
\Omega_{r,0}\left(\frac{a_0}{a}\right)^4 +
\Omega_{m,0}\left(\frac{a_0}{a}\right)^3 +
\Omega_{\Lambda,0}
}\,,\label{eq:H0EvoStandard}
\end{align}
where 
$H_0 = 100\,h\,\textrm{km}\,\textrm{s}^{-1}\, \textrm{Mpc}^{-1}$ is the 
present-day Hubble expansion rate while $\Omega_{i,0}\equiv \rho_{i,0}/\rho_{\textrm{crit},0}$ are the dimensionless density parameters with $i=r,m,\Lambda$ denoting, respectively, radiation, matter and vacuum energy. The critical density of the universe today is given by $\rho_{\textrm{crit},0} = 3H_0^2/8\pi G_N$ with $G_N$ the Newtonian constant of gravitation. 

\begin{figure}[!t]
	\centering
\includegraphics[width=0.495\textwidth]{ 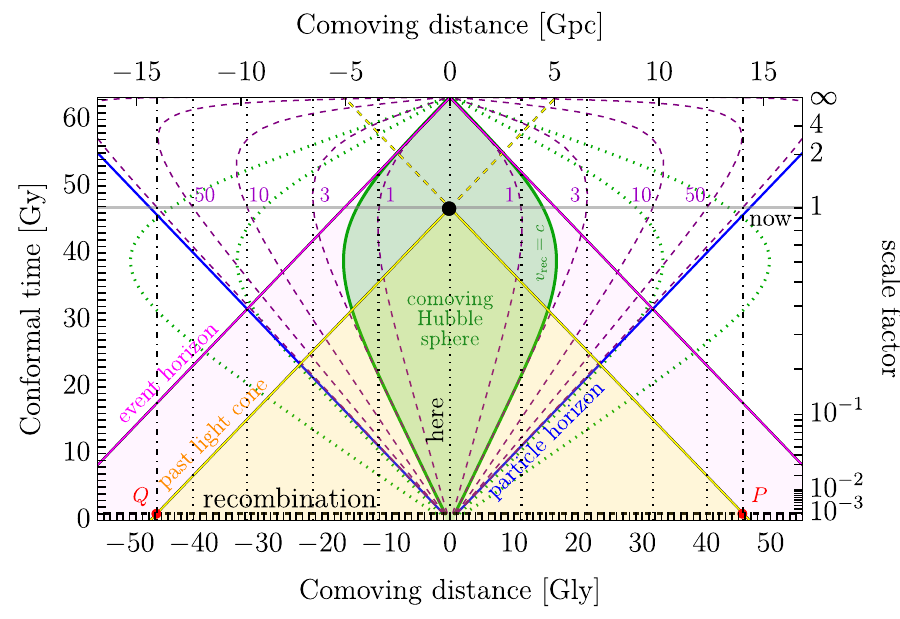}
	\caption{ \it 
Spacetime diagram of the Universe, cf. the text in section\,\ref{sec:ScalesUniverse} for details. 
 }
\label{fig:HorizonProblem}
\end{figure}

After integration, we get
\begin{align}
\tau(a)-\tau_{\textrm{in}} = 
\int_{a_{\textrm{in}}}^{a}
\frac{d a^{\prime}}{a^{\prime 2}
H(a^{\prime})}\,.\label{eq:ConformalTimeInte}
\end{align}
In the standard cosmological evolution without inflation, we integrate from the initial big bang singularity at $\tau_{\textrm{in}} = 0$ with $a_{\textrm{in}} = 0$. 
Through Eq.\,(\ref{eq:ConformalTimeInte}), it is possible, for a given value of the scale factor $a$ and thus for a fixed value of the comoving Hubble radius, to compute the corresponding conformal time. 
The result is shown in Fig.\,\ref{fig:HorizonProblem}. 
The two solid green lines describe the evolution of $\pm(aH)^{-1}$ as the scale factor $a$ and therefore the changes of conformal time.
To draw this plot, we consider $\theta = \phi = \textrm{const}$ and the comoving distance is, therefore, purely radial.
Our position in today's universe corresponds to the black dot at vanishing comoving distance (``here'') and conformal time $\tau_{\textrm{now}} \simeq 46.8$ Gy (``now''). 
At fixed $\theta$ and $\phi$, positive and negative comoving distances indicate two opposite directions in the sky. 
The green shaded region, therefore, represents the so-called comoving Hubble sphere. For each $\tau$, this is the comoving region encompassed by $[-R_H,+R_H]$. 
The comoving Hubble sphere expands or contracts depending on the dominant energy content of the universe. During the radiation and matter-dominated eras, it expands, while during the accelerated expansion of the dark energy-dominated era, it contracts. 
This fact imparts a peculiar teardrop shape to the evolution of the comoving Hubble radius.
From Hubble’s law, it follows that the comoving Hubble sphere marks the boundary where the recession velocity $v_{\textrm{rec}}$ of galaxies is less than the speed of light in comoving coordinates.  
This sphere, therefore, represents a significant threshold in the universe since light emitted now from galaxies within the comoving Hubble sphere can eventually reach us, making them observable in principle. 
To illustrate this point, we write the recession velocity $v_{\textrm{rec}}(t,z)$ as $v_{\textrm{rec}}(t,z) = 
 \dot{a}(t)\chi(z) = \chi(z)/R_H(t)$,
where $\chi(z)$ is the fixed comoving coordinate associated with a galaxy observed today at redshift $z$. 
The relation between redshift $z$ and scale factor is $1+z = a_0/a(t)$, and $\chi(z)$ can be computed according to the distance-redshift relation
$\chi(z) = \frac{1}{a_0}\int_{0}^{z}\frac{dz^{\prime}}{H(z^{\prime})}$.
In Fig.\,\ref{fig:HorizonProblem}  the vertical dotted lines show the worldlines of comoving objects at comoving distances fixed by $\chi(z)$ with $z=1,3,10,50$. 
Each fixed comoving coordinate, therefore, can be thought of as a galaxy observed today at redshift $z$ (indicated with a purple number along the ``now'' line). 
It is now mathematically clear that the condition $\chi(z) > R_H(t_0)$ corresponds to recession velocities exceeding the speed of light and to comoving distances beyond the comoving Hubble sphere. 
For illustration, the dotted green line corresponds to recession velocities equal to twice and three times the speed of light in comoving coordinates.
Note that the redshift of an object at fixed comoving coordinate changes with time. In Fig.\,\ref{fig:HorizonProblem}, this is illustrated by the dashed purple lines that indicate constant values of redshift.

The yellow-shaded region (labeled past light cone) encompasses all events in the universe that are observable to us, extending up to the present moment. Its boundary, defined by the equation
$\chi_{\textrm{lc}}(t_{\textrm{em}}) =     \int_{t_{\textrm{em}}}^{t_0}
\frac{dt^{\prime}}{a(t^{\prime})}$,
includes all celestial locations from which light is currently reaching us.

The lightly shaded magenta area labeled as the event horizon encapsulates everything observable from our current location throughout the entire chronology of the universe, extending indefinitely into the future. Objects and events outside this region, 
bounded by the equation
  $\chi_{\textrm{eh}}(t) = 
  \int_t^{t_{\textrm{end}}}
  \frac{dt^{\prime}}{a(t^{\prime})}$,
will remain forever invisible to us and will have no impact on our observations.

The particle horizon, represented by the blue lines corresponding to 
$\chi_{\textrm{ph}}(t) =    
  \int_{0}^{t}
  \frac{dt^{\prime}}{a(t^{\prime})}$,
denotes the distance light could have traveled since the initial time, defining the extent of the observable universe at any given moment. It is notable that the interval marked by the intersection of the present moment line and the blue lines corresponds to the width of the past light cone depicted at the diagram's base. Additionally, we observe that the blue lines correspond to regions associated with infinite redshift.

Finally, for completeness, in Fig.\,\ref{fig:HorizonProblem} we also show, in addition to the past, the future of the universe, taking into account the present accelerating expansion driven by dark energy. Notice that the comoving Hubble sphere shrinks to a point.
This phenomenon is a direct consequence of the accelerating expansion of the universe: as the rate of expansion increases, galaxies that are already receding from us will do so at an accelerating rate. Eventually, this will lead to a situation where regions of space not gravitationally bound to us, such as distant galaxies, will move beyond our observable horizon, effectively creating an `island universe' confined to our local gravitationally bound structures.
In conclusion, inside the comoving Hubble radius, regions
are causally connected, meaning that they can influence
each other via light signals or other causal interactions, while outside this radius they are causally disconnected.

The horizontal dashed line at the bottom of the plot in Fig.\,\ref{fig:HorizonProblem} indicates the recombination time, representing the epoch in the history of the universe when electrons and protons combined to form neutral hydrogen atoms. Shortly after recombination, photon decoupling occurred---an epoch when photons became free to travel through space. Before photon decoupling, photons were tightly coupled to charged particles in the ionized plasma, constantly scattering and preventing light from freely propagating through space. After recombination, photons were able to decouple and travel freely, giving rise to the CMB radiation we observe today. Recombination and photon decoupling occurred around the same redshift, approximately at $z_{\textrm{rec}} = 1090$. 

\begin{figure}[!t]
	\centering
\includegraphics[width=0.495\textwidth]{ 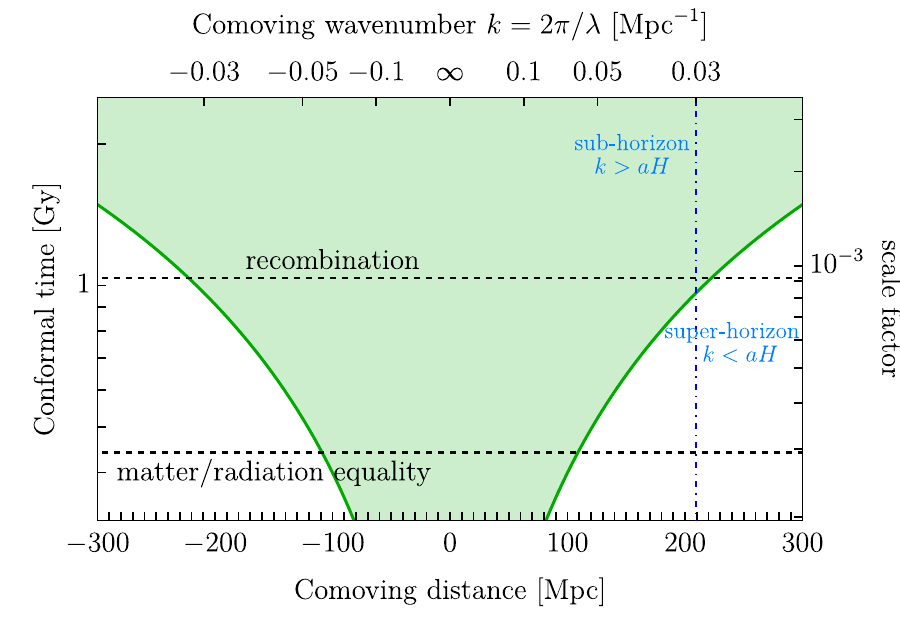}
	\caption{\it Evolution of the comoving distance (x-axis) in terms of the conformal time (y-axis). The green region shows the distance inside which, at a given time, two points in the universe are causally connected.
 }
\label{fig:HorizonProblemZoom}
\end{figure}

In Fig.\,\ref{fig:HorizonProblemZoom}, we zoom in on the evolution of the comoving Hubble radius during the period of recombination.

Notice that we changed the units of comoving distances from Gly to Mpc. 
This zoom allows us to highlight the period of matter-radiation equality---that is, the epoch in the cosmological history of the universe when the energy density contributed by matter becomes equal to the energy density contributed by radiation---occurring at a redshift $z_{\textrm{eq}} = 3402$.

On the upper $x$-axis, we indicate, for each value of the comoving distance $\lambda$, the corresponding comoving wavenumber calculated as $k=2\pi/\lambda$. The main point of this analysis is to emphasize that small scales characterized by $k\gtrsim 0.05$ Mpc$^{-1}$ lie within the comoving Hubble horizon at the time of matter-radiation equality. Physically, these small scales are in causal contact at the time of matter-radiation equality, and their subsequent evolution under the influence of gravity seeds the large-scale structure of the observed universe. After matter-radiation equality, the universe became matter-dominated, allowing perturbations to grow via gravitational instability, leading to the formation of the structures we see today.

The horizon problem, solved by inflation, is showed by the conformal diagram in Fig.\,\ref{fig:HorizonProblem}. 
Imagine two diametrically opposed directions in the sky. The CMB photons reaching us from these directions originated at points labeled $P$ and $Q$ in Fig.\,\ref{fig:HorizonProblem}. Notably, these photons were emitted in such proximity to the big bang singularity that the past light cones of $P$ and $Q$ do not intersect. 
The puzzle lies in understanding how the photons originating from points $P$ and $Q$ possess nearly identical temperatures without any possibility of direct communication between them.
What happens, with the introduction of an inflationary phase is that the big bang singularity at $\tau = 0$ is formally extended to $\tau = -\infty$.
The radiation era is preceded by an inflationary phase dominated by the dynamics of the inflaton field during which there is a shrinking of the comoving Hubble sphere. 
An immediate consequence of this setup is evident by examining Fig.\,\ref{fig:HorizonProblemZoom}. 
If we imagine extending the lower part of the plot with a phase in which there is a shrinking of the comoving Hubble radius, what happens is that a certain value of $k$, which exits the comoving Hubble sphere during the radiation era (tracing back in time), will re-enter it at some point during the inflationary phase if it persists for a sufficiently long duration. As an illustrative example, consider the value $k=0.03$ Mpc$^{-1}$ (vertical dot-dashed blue line), for which we have $k>aH$ after recombination (sub-horizon mode) and $k<aH$ before recombination (super-horizon mode). This value of $k$ will re-enter the sub-horizon regime at some point during inflation (for a better and more immediate understanding of this dynamics, it is possible to take a sneak peek in advance at Fig.\,\ref{fig:StaroPlot}).

In order to derive the master formula in eq.\,\ref{eq:MainLog}, we start from
\begin{align}
\log\left(
\frac{k}{a_{\textrm{eq}}H_{\textrm{eq}}}
\right) &= 
-\Delta N_k - \Delta N_{\textrm{reh}} \nonumber \\  &- \Delta N_{\textrm{RD}} + \log\left(
\frac{H_k}{H_{\textrm{eq}}}
\right)\,.\label{eq:Master1}
\end{align}
As previously mentioned, we characterize the reheating period as a time when the universe is governed by a fluid exhibiting a constant equation of state, denoted as $\omega_{\textrm{reh}}$. Consequently, the continuity equation $\dot{\rho} + 3H\rho(1+\omega_{\textrm{reh}}) = 0$ reads 
$d\log\rho = -3(1+\omega_{\textrm{reh}})d\log a$ from which, using Eq.\,(\ref{eq:EfoldNumber}), we get 
\begin{align}
\rho(N_{\textrm{reh}}) = \rho(N_{\textrm{end}}) 
e^{-3(1+\omega_{\textrm{reh}})(N_{\textrm{reh}}-N_{\textrm{end}})}\,,\label{eq:reheatingdensity}
\end{align}
where we integrated between the end of inflation and the end of the reheating phase (cf. point {\it ii)} above). 
In Eq.\,(\ref{eq:reheatingdensity}), 
$\rho(N_{\textrm{end}}) \equiv \rho_{\textrm{end}}$ is the energy density at the end of inflation, and can be computed for a given explicit model of inflation. 
On the other hand, $\rho(N_{\textrm{reh}})\equiv \rho_{\textrm{reh}}$ is the energy density at the end of the reheating phase. 
The latter can be related to the reheating temperature of the universe $T_{\textrm{reh}}$ through 
\begin{align}
\rho_{\textrm{reh}} = \left(
\frac{\pi^2}{30}
\right)g_{\textrm{reh}}T_{\textrm{reh}}^4\,.  
\end{align}
The conservation of comoving entropy gives a relation between the reheating temperature and the present-day temperature $T_0$. 
Consequently, Eq.\,(\ref{eq:reheatingdensity}) reads
\begin{align}
\frac{3(1+\omega_{\textrm{reh}})}{4}\Delta N_{\textrm{reh}} & = \frac{1}{4}\log\left(
\frac{30}{g_{\textrm{reh}}\pi^2}
\right) + 
\frac{1}{4}\log\left(
\frac{\rho_{\textrm{end}}}{T_0^4}
\right)\nn\\
& +\frac{1}{3}\log\left(
\frac{11g_{s,\textrm{reh}}}{43}
\right) + \log\left(
\frac{a_{\textrm{eq}}}{a_0}
\right)\nn\\
&- \Delta N_{\textrm{RD}}\,,\label{eq:Master2}
\end{align}
Both $g_{s,\textrm{reh}}$ and $g_{\textrm{reh}}$ depends on temperature. 
Since they enter only logarithmically, for the sake of simplicity, we choose to fix their values to the fiducial value $g_{s,\textrm{reh}} = g_{\textrm{reh}} = 100$.
We use Eq.\,(\ref{eq:Master2}) to eliminate the 
$\Delta N_{\textrm{RD}}$-dependence in  Eq.\,(\ref{eq:Master1}). 
We thus arrive at 
\begin{align}
\log\left(
\frac{k}{a_{\textrm{eq}}H_{\textrm{eq}}}
\right) & = -\Delta N_{k} +\frac{(3\omega_{\textrm{reh}}-1)}{4}\Delta N_{\textrm{reh}} \nn\\ & 
-
\frac{1}{4}\log\left(
\frac{30}{g_{\textrm{reh}}\pi^2}
\right)-
\frac{1}{3}\log\left(
\frac{11g_{s,\textrm{reh}}}{43}
\right) \nn\\ &
-\log\left(
\frac{a_{\textrm{eq}}}{a_0}
\right)\underbrace{-\log\left(
\frac{\rho_{\textrm{end}}^{1/4}}{T_0}
\right) + \log\left(
\frac{H_k}{H_{\textrm{eq}}}
\right)}_{=\,\frac{1}{2}\log\left(
\frac{H_k}{\sqrt{3}\MPl}
\right) + \log\left(
\frac{T_0}{H_{\textrm{eq}}}
\right)}\,.
\end{align}
Since $\rho_{\textrm{end}}=3\MPl^2H_{\textrm{end}}^2$, with $H_{\rm end}\simeq H_k$ almost constant during inflation, we arrive at the final equation 
\begin{align}
\log\left(
\frac{k}{a_{\textrm{eq}}H_{\textrm{eq}}}
\right) & = -\Delta N_{k} + \frac{1}{2}
\log\left(\frac{H_k}{\sqrt{3}\MPl}\right)
+ \log\left(\frac{T_0}{H_{\textrm{eq}}}\right)\nn\\
& +\frac{(3\omega_{\textrm{reh}}-1)}{4}\Delta N_{\textrm{reh}} -
\frac{1}{4}\log\left(
\frac{30}{g_{\textrm{reh}}\pi^2}
\right)
\nn\\
& -
\frac{1}{3}\log\left(
\frac{11g_{s,\textrm{reh}}}{43}
\right) - \log\left(\frac{a_{\textrm{eq}}}{a_0}\right)\,.\label{eq:MainLog2}
\end{align}.

\vspace{-1cm}
\section{Starobinsky model in slow-roll approximation}\label{app:Staro}

In this appendix, we present the semi-analytical expressions corresponding to the main quantities that characterize the Starobinsky inflationary model in the slow-roll approximation. 
The model is based on the scalar potential
\begin{align}
V_{\textrm{Staro}}(\phi) = 
\frac{3M^2\MPl^2}{4}
\left[1 - 
\exp\left(
-\sqrt{\frac{2}{3}}
\frac{\phi}{\MPl}
\right)
\right]^2\,.
\end{align}
For ease of reading, we introduce the function
\begin{align}
f_x \equiv -\frac{1}{3}
(3+2\sqrt{3})e^{-1-2/\sqrt{3} - 4x/3}\,.\label{eq:ShortHandfx}
\end{align}
We indicate with $\Delta N_{\star}$ the number of $e$-folds between the horizon crossing of the CMB pivot scale $k_{\star} \equiv 0.05$ Mpc$^{-1}$ and the end of inflation.
The field value at the end of inflation is
\begin{align}
\phi_{\textrm{end}} = 
\sqrt{\frac{3}{2}}
\log\left(1+\frac{2}{\sqrt{3}}\right)\MPl\,.
\end{align}
The field value at the CMB pivot  scale is
\begin{align}
\phi_{\textrm{CMB}} = 
-\frac{1}{\sqrt{6}}\big[&
3+2\sqrt{3}+4
\Delta N_{\star} 
+ 
\log(-135+78\sqrt{3})
\nn \\ & + 3W_{-1}(f_{\Delta N_{\star}})\big]
\MPl\,,
\end{align}
where $W_{-1}(z)$ is the branch with 
$k=-1$ of the Lambert W function $W_k(z)$. 
The scalar spectral index at the CMB pivot scale reads
\begin{align}
n_s = 1-\frac{16}{3[1+W_{-1}(f_{\Delta N_{\star}})]^2} 
+ \frac{8}{
3[1+W_{-1}(f_{\Delta N_{\star}})]
}\,,
\end{align}
while for the tensor-to-scalar ratio we find
\begin{align}
r = 
\frac{64}{
3[1+W_{-1}(f_{\Delta N_{\star}})]^2
}\,.
\end{align}
The amplitude of the scalar power spectrum at the CMB pivot scale is
\begin{align}
A_s = \frac{3M^2[1+W_{-1}(f_{\Delta N_{\star}})]^4}{128\pi^2 \MPl^2 
W_{-1}(f_{\Delta N_{\star}})^2}\,.
\end{align}
Finally, the square of the Hubble rate at the end of inflation is 
\begin{align}
H(N_{\textrm{end}})^2\equiv H_{\textrm{end}}^2 =  
3\left(\frac{7}{2}-2\sqrt{3}\right)M^2\,,
\end{align}
while its value at the time of horizon crossing for the CMB pivot scale  $k_{\star}$ is
\begin{align}
H(N_{\star})^2\equiv
H_{\star}^2 = \frac{M^2}{4}
\left[
1+\frac{1}{W_{-1}(f_{\Delta N_{\star}})}
\right]^2\,.
\end{align}
The value of the Hubble rate at the $e$-fold time $N_k$ at which the generic comoving wavenumber $k$ crosses the inverse comoving Hubble radius (that is, the time $N_k$ defined by the condition $k=a(N_k)H(N_k)$) can be obtained from the scaling
\begin{align}
\frac{k}{k_{\star}}=\frac{
a(N_k)H(N_k)
}{
a(N_{\star})H(N_{\star})
}\to
H(N_k)\equiv H_k = 
\left(\frac{k}{k_{\star}}\right)e^{N_* - N_k}H_{\star}\,.
\end{align}

\section{Details of the scalar potential}\label{app:Para}
Here we report briefly the relation between the parameter in the different potential parameterizations.
\begin{align}
&\alpha_2  =\nn\\
&\frac{4 x_0^3 [x_1^2 (\beta_2 - 2 \beta_3) - 2 x_0 x_1 (\beta_3 + \beta_5) +
    x_0^2 (-2 \beta_5 + \beta_6)]}{
x_0^2 + 4 x_0 x_1 + x_1^2
    }\,,\\
&-\alpha_3  =\nn\\
& \frac{4 x_0^2 [-x_1^2 (\beta_2 - 4 \beta_3) + 
   4 x_0 x_1 (\beta_3 + 2 \beta_5) + x_0^2 (8 \beta_5 - 5 \beta_6)]}{
   x_0^2 + 4 x_0 x_1 + x_1^2
   }\,,\\
&\alpha_5  =\nn\\
&\frac{4 x_1^3 [x_0^2 (\beta_2 - 2 \beta_3) - 2 x_0 x_1 (\beta_3 + \beta_5) +
    x_1^2 (-2 \beta_5 + \beta_6)]}{
x_0^2 + 4 x_0 x_1 + x_1^2    
    }\,,\\
&\alpha_6  =\nn\\
& \frac{4 x_1^2 [x_0^2 (\beta_2 - 4 \beta_3) - 
   4 x_0 x_1 (\beta_3 + 2 \beta_5) + x_1^2 (-8 \beta_5 + 5 \beta_6)]}{
 x_0^2 + 4 x_0 x_1 + x_1^2  
   }\,.
\end{align}
\section{The PBH abundance}\label{app:PBHabu}
The fraction of energy density $\beta_k(M_{\rm PBH}) \td  \ln M_{\rm PBH}$ collapsing into PBHs can be estimated as 
\be\label{eq:betak}
    \beta_k(M_{\rm PBH})
    = \int_{\mathcal{C}_{\rm th}} \! \td\mathcal{C} \, P_k(\mathcal{C}) \frac{M_{\rm PBH}}{M_k}  \delta\left[ \ln\frac{M_{\rm PBH}}{M_{\rm PBH}(\mathcal{C})} \right]\!,
\ee
where $P_k(\mathcal{C})$ denotes the probability a black hole will form in the Hubble patch. The PBH mass function can be obtained directly from the collapse fraction:
\be\label{eq:df_PBH}
    f_{\rm PBH}(M_{\rm PBH})
= \frac{1}{\Omega_{\rm DM}}\int \frac{\td M_k}{M_k} \, \beta_k(M_{\rm PBH} ) \left(\frac{M_{\rm eq}}{M_k}\right)^{1/2} \!\!,
\ee
where $M_{\rm eq} \approx 2.8\times 10^{17}\,\,M_{\odot}$ is the horizon mass at the time of matter-radiation equality and $\Omega_{\rm  DM} = 0.12h^{-2}$ is the cold dark matter density~\cite{Planck:2018jri}.

The PBH mass function depends on the formalism used. The two most common approaches in the literature are threshold statistics on the compaction function (used in the main text) and peaks theory\footnote{Since the amount of primordial non-gaussianities is small for USR models\,\cite{Atal:2018neu,Franciolini:2022pav,Firouzjahi:2023xke,Frosina:2023nxu}, here we safely neglect their impact in the computation of the abundance\,\cite{Young:2022phe,Ferrante:2022mui,Gow:2022jfb,Ianniccari:2024bkh}.}.

\textit{Threshold statistics} (TS)-- In the main text of this work, we used the threshold statistics formalism where the probability can be estimated from the statistics of the compaction function $\mathcal{C}$~\cite{Ferrante:2022mui,Gow:2022jfb}, generically defined as twice the local mass excess over the areal radius.
Within this formalism, the mass function can be computed as\,\cite{Young:2019yug,DeLuca:2019qsy}
\begin{equation}
\begin{aligned}
&f_{\rm PBH}(M_{\rm PBH}) = \frac{1}{\Omega_{\rm DM}}\int_{M_H^{\rm min}}^{\infty}\frac{\td M_H}{M_H}
\left(
\frac{M_{\rm eq}}{M_H}
\right)^{1/2}\left(
\frac{M_{\rm PBH}}{\gamma M_H}
\right) \\& \times
\left(\frac{M_{\rm PBH}}{\mathcal{K}M_H}
\right)^{1/\gamma}\!\!\frac{
1
}{
\sqrt{2\pi}\sigma_c(M_H)
\Lambda^{1/2}}\textrm{exp}\Bigg[-\frac{8\left(1 - 
\sqrt{\Lambda}\right)^2}{9\sigma_c^2(M_H)}\Bigg]
\,,
\end{aligned}
\end{equation}
where 
\begin{equation}
\Lambda=1 - \left({\cal C}_{\rm th} - \frac{3\left(M_{\rm PBH}/(\mathcal{K}M_H)\right)^{1/\gamma}}{2}\right)
\end{equation}
and the lower limit of integration follows from the condition $\Lambda > 0$.

The variance can be computed as
\begin{equation}
    \sigma_c^2(M_H) = \frac{16}{81}
\int_{0}^{\infty}\frac{\td k}{k}(k r_m)^4W(k r_m)^2 P_{\zeta}^{\mathcal{T}}(k,r_m)\label{eq:Sigma2}
\end{equation}
with $P_\zeta^T=T^2\left(k, r_m\right) P_\zeta(k)$. We have defined $W\left(k, r_m\right)$ and $T\left(k, r_m\right)$ as the top-hat window function and the radiation transfer function\,\cite{Young:2022phe}.

The parameters $\gamma$, $r_m$, $\mathcal{K}$ and $C_{\rm th}$ depend on the shape of the power spectrum\,\cite{Musco:2020jjb,Byrnes:2018clq,Musco:2023dak,Ianniccari:2024ltb}, and in this work we use $\gamma=0.36$, $C_{\rm th}=0.54$, $r_m=3.4$ and $\mathcal{K}=5$.
We neglect for simplicity the QCD impact on these parameters\,\cite{Musco:2023dak}.

\textit{Peak theory} (PT)-- This alternative formalism identifies PBHs with sufficiently high peaks in the overdensity field\,\cite{Yoo:2018kvb,Yoo:2019pma,Franciolini:2022tfm}. In the high peaks limit the mass function can be computed as 
\begin{equation}
\begin{aligned}
&f_{\mathrm{PBH}}\left(M_{\mathrm{PBH}}\right)= \frac{1}{\Omega_{\mathrm{DM}}} \int_{M_{\mathrm{H}}^{\min }} \frac{d M_{\mathrm{H}}}{M_{\mathrm{H}}}\left(\frac{M_{\mathrm{eq}}}{M_{\mathrm{H}}}\right)^{1 / 2}\\
& \times \left(\frac{M_{\mathrm{PBH}}}{\mathcal{K} M_{\mathrm{H}}}\right)^{\frac{1+\gamma}{\gamma}} \frac{\mathcal{K}}{\gamma} \left(\frac{2}{3}\right)^4\frac{(1-\sqrt{\Lambda}))^3}{ \pi \sigma_c^4(M_H) \Lambda^{1 / 2}} \left(\frac{\sigma_{c c}(M_H)}{\sigma_c(M_H)}\right)^3 \\ & \times \textrm{exp}\Bigg[-\frac{8\left(1 - 
\sqrt{\Lambda}\right)^2}{9\sigma_c^2(M_H)}\Bigg]\,
\end{aligned}
\end{equation}
where the first dimensionless rescaled moment of the distribution~\cite{Young:2014ana,Young:2020xmk,Musco:2023dak}
\begin{equation}
\begin{aligned}
    \sigma_{cc}^2(M_H) =(r_m \sigma_1)^2 & = \frac{16}{81} \int_0^{\infty} \frac{\td \ k} {k}\left(k r_m\right)^6  \\ & \times W^2\left(k, r_m\right)P_{\zeta}^{\mathcal{T}}(k,r_m)\,.
\end{aligned}
\end{equation}

It is well known that the threshold statistics formalism does not agree with the theory of peaks (see, e.g., Refs.~\cite{Green:2004wb, Young:2014ana, DeLuca:2019qsy,Iovino:2024tyg}), with the second approach tending to overproduce PBHs respect the first one. Since the abundance is exponentially sensitive to the amplitude of the power spectrum, this discrepancy is reflected by a small fine-tuning of the amplitude of the power spectrum to get the same mass function in the two approaches. To get the same mass function within the peak theory, as depicted in Fig.\,\ref{fig:PBHAbu},it is necessary to decrease the amplitude of the power spectrum at the peak from $A=10^{-1.97}$ to $A=10^{-2.07}$. This is possible with a very small decrease of the potential parameter $\delta\beta_3=\beta^{\rm TS}_3-\beta^{PT}_3\lesssim10^{-8}$ while keeping the others fixed. In this way we were able to find the same mass function in the two approaches with the same inflationary parameters ($n_s,r$) and a negligible change in the total number of efolds ($\delta N =N_{\rm TS}-N_{\rm PT}\lesssim 0.5$). Hence, the claim presented in this paper is valid within both approaches.
\bibliography{main}

\end{document}